\documentclass[a4paper,american,english]{article}
\usepackage[T1]{fontenc}
\usepackage[latin9]{inputenc}
\usepackage{color}
\usepackage{babel}
\usepackage{verbatim}
\usepackage{url}
\usepackage{amsmath}
\usepackage{amssymb}
\usepackage{cancel}
\usepackage{graphicx}
\PassOptionsToPackage{normalem}{ulem}
\usepackage{ulem}

\makeatletter

\providecommand{\tabularnewline}{\\}

\newcommand{\lyxaddress}[1]{
    \par {\raggedright #1
    \vspace{1.4em}
    \noindent\par}
}

\usepackage{huncial}
\usepackage{xcolor}

\def\ABK{\texthuncl{A}}

\usepackage{babel}

\makeatother

\begin{document}
\title{Charged pion decay in a laser field}
\author{D. Gomez Dumm$^{a}$, S. Noguera$^{b}$ and N.N.\ Scoccola$^{c,d}$}
\date{}
\maketitle

\lyxaddress{$^{a}$ IFLP, CONICET $-$ Departamento de F\'isica, Facultad de Ciencias Exactas,
Universidad Nacional de La Plata, C.C. 67, 1900 La Plata, Argentina}

\lyxaddress{$^{b}$ Departamento de F\'isica Te\'orica and IFIC, Centro Mixto Universidad
de Valencia-CSIC, E-46100 Burjassot (Valencia), Spain}

\lyxaddress{$^{c}$ CONICET, Rivadavia 1917, 1033 Buenos Aires, Argentina}

\lyxaddress{$^{d}$ Physics Department, Comisi\'on Nacional de Energ\'ia
At\'omica, Av.\ Libertador 8250, 1429 Buenos Aires, Argentina\vspace*{0.5cm}
 }
\begin{abstract}
We study the decays $\pi^{+}\to l^{+}\nu_{l}$ in the presence of a
background electromagnetic plane wave with circular polarization. We find
that, in the presence of this background, the number of hadronic form
factors required to describe the pion-to-vacuum amplitude increases from one
to four. Three of these form factors are associated with the axial vector
current and one with the vector piece, which is in general different from
zero. We obtain analytical expressions for the corresponding contributions
to the decay width, considering in particular some range of values of
interest for the frequency and amplitude of the external wave. For laser
frequencies $\omega$ of the order of the eV, it is seen that the main
correction to the result obtained in absence of the external field is
proportional to the vector form factor, carrying a suppression factor of
order $\sim(\omega/m_{\pi})^{4/3}$. In addition, we show that our
expressions can be connected with the Kroll-Watson formula, which provides
the differential cross section for the scattering of a charged particle by
some potential in the presence of a strong external electromagnetic wave.
Our analytical results are also compatible with previous numerical
calculations obtained in the limit of low amplitude laser fields, offering
new insight on some characteristics of these results. \vspace{2cm}
\foreignlanguage{american}{}
\end{abstract}
\selectlanguage{american}%
\selectlanguage{english}%

\section{Introduction}

The recent progress in laser technology, in terms of both its intensity and
its sources, is paving the way for the experimental study of quantum effects
in the presence of a very large electromagnetic wave
field~\cite{Fedotov:2022ely,DiPiazza:2011tq,Seipt:2017ckc}. From the
theoretical point of view this subject has a rather long history. For
example, already in the mid 1960s, the effect of some electromagnetic field
configurations on leptonic decays of charged pions was
considered~\cite{Nikishov:1964zza,Nikishov:1964zz,Ritus1969}. Over the
years, several other decay processes of interest in nuclear and particle
physics like e.g.\ the muon weak decay~\cite{Ritus:1985vta}, the $\beta$
decay of nuclei~\cite{Akhmedov:1983gts,Nikishov:1983abc}, the
transmutation of protons~\cite{Wistisen:2020czu}, etc., have been also
analyzed  in this context. More recent applications, as well as related
references, can be found in Ref.~\cite{Fedotov:2022ely}. Whether noticeable
effects on charged particle weak decays may be observed in present or
forthcoming experimental facilities is an issue that shows some controversy.
For example, it has been claimed~\cite{Liu:2007zzg} that the muon lifetime
can change dramatically in the presence of laser fields having an electric
field amplitude of $10^{6}$ V/cm. This result has been questioned in
Ref.~\cite{Narozhny:2008zz} stating that such a field is in fact too weak to
produce an observable effect, and this argument has been confirmed in a more
quantitative way in Refs.~\cite{Dicus:2008nw,Farzinnia:2009gg}.

Due in part to these developments, in Ref.~\cite{Mouslih:2020pfd} the weak
decay of charged pions has been revisited. In that article, as in
Ref.~\cite{Nikishov:1964zz}, the authors write the pion-to-vacuum hadronic
current in terms of a single form factor (the usual pion decay constant in
vacuum), and consider laser fields with electric field amplitudes between
$10^{7}-10^{8}$~V/cm and frequencies $\omega=1-2$~eV. They conclude that
within this range of laser parameters the calculated decay width basically
coincides with the one obtained in vacuum. One important observation is that
to get this result one has to sum over the contributions of a rather large
number of exchanged photons. Since in Ref.~\cite{Mouslih:2020pfd} such a sum
is performed numerically, the analysis of the process for higher electric
field amplitudes lies beyond their computing capacity. Forthcoming
experimental facilities like e.g.\
LUXE~\cite{LUXE:2023crk,Abramowicz:2021zja} or NSF-OPAL~\cite{nsfopal} are,
however, expected to reach electric field amplitudes about 5 orders of
magnitude larger, with about the same laser frequency. Thus, one might
wonder whether the conclusions of Ref.~\cite{Mouslih:2020pfd} would remain
valid for these parameters. In addition, as shown in
Ref.~\cite{Coppola:2018ygv}, the presence of a background electromagnetic
field opens the possibility of the existence of other nonzero pion-to-vacuum
transition form factors in addition to the pion decay constant considered in
Refs.~\cite{Nikishov:1964zz,Mouslih:2020pfd}. These form factors have
been shown to play an important role in the analysis of the weak decay width
of charged pions in the presence of a constant magnetic
field~\cite{Coppola:2019idh}. Given this context, the scope of our work is
twofold. Firstly, considering the presence of an external plane wave (PW)
electromagnetic field, we determine the form of the most general
model-independent hadronic matrix elements, written in terms of gauge
covariant Lorentz structures. In particular, we show that in the presence of
a PW one can in principle define four independent form factors. Secondly,
given the general form of the hadronic matrix elements, we obtain a general
expression for the charged pion decay width  and show that this expression
can be conveniently approximated so as to obtain reliable numerical results
for relevant modern laser parameters.

This article is organized as follows. In Sec.\ 2 we review $\pi^{+}\to
l^{+}\nu_{l}$ decays in vacuum and discuss the main features of the external
field. Then, in Secs.\ 3 and 4 we obtain general expressions for the
relevant matrix elements and for the corresponding decay width,
respectively. In Sec.\ 5 we derive approximate analytical expressions for
the contributions to the decay width, taking into account relevant parameter
ranges for the laser field. The order of magnitude of these contributions is
discussed, and it is shown that in the limit of low amplitude laser fields
the results are compatible with previous numerical calculations. It is also
seen that our analytical results can be related to the Kroll-Watson formula.
Finally, in Sec.\ 6 we present a summary of our work, together with our main
conclusions. Specifications about the characteristics of pion and lepton
fields in the presence of the external background are included in App.\ A,
while in App.\ B we detail some mathematical steps followed in our
calculations.

\section{Preliminaries}

\newcounter{eraI}
\renewcommand{\thesection}{\arabic{eraI}}
\renewcommand{\theequation}{\arabic{eraI}.\arabic{equation}}
\setcounter{eraI}{2} \setcounter{equation}{0} 

We are interested in the two-body decays of a charged pion into a charged
lepton (electron or muon) and the corresponding neutrino, in the presence of
an electromagnetic plane wave background field that can be understood as an
idealization of a laser field. In particular, for simplicity, we consider
the case in which the electromagnetic wave is circularly polarized and the
pion moves in the same direction as the laser beam. We use natural
Heaviside-Lorentz units with $\hbar=c=1$ and fine structure constant
$\alpha=e^{2}/4\pi\simeq1/137.$

\subsection{$\pi^{\pm}\to l^{\pm}\nu_{l}$ decays in vacuum}

Let us start by analyzing leptonic pion decays in absence of the external
field, i.e., in vacuum. For concreteness we consider the process
$\pi^{+}(q)\rightarrow l^{+}(p)\,\nu_{l}(t)$, where $l=e,\mu$, denoting by
$q$, $p$ and $t$ the four-momenta of the pion, charged lepton and neutrino,
respectively.

The decay is driven by the Fermi Lagrangian density
\begin{eqnarray}
{\cal L}_{\text{F}}(x) & = & -\frac{G_{F}}{\sqrt{2}}\,V_{\text{ud}}\,\sum_{l}\left[J_{h}^{\dagger\mu}(x)\,J_{l\mu}(x)+J_{h}^{\mu}(x)\,J_{l\mu}^{\dagger}(x)\right]\ ,\label{FermiLagran}
\end{eqnarray}
where $G_{F}$ is the Fermi decay constant, $V_{\text{ud}}$ is the
corresponding element of the Cabibbo-Kobayashi-Maskawa matrix and
$J_{l}^{\dagger\mu}(x)$, $J_{h}^{\mu}(x)$ are, respectively, the
leptonic and hadronic currents involved in the $\pi^{+}(q)\rightarrow l^{+}(p)\,\nu_{l}(t)$
process. These currents are given by
\begin{equation}
J_{l}^{\dagger\mu}(x)=\bar{\psi}_{\nu_{l}}(x)\gamma^{\mu}(1-\gamma^{5})\psi_{l}(x)\ ,
\qquad\qquad J_{h}^{\mu}(x)=\bar{\psi}_{d}(x)\gamma^{\mu}(1-\gamma^{5})\psi_{u}(x)\ .
\label{currents}
\end{equation}
In vacuum, the decay amplitude can be written in terms of the matrix
elements
\begin{equation}
\left\langle \bar{l}(p,s)\,\nu_{l}(t,r)\right|\bar{\psi}_{\nu_{l}}(x)\gamma^{\mu}(1-\gamma^{5})\psi_{l}(x)\left|0\right\rangle \ =\ e^{it\cdot x}\,\bar{u}_{\nu_{l}}(t,r)\,\gamma^{\mu}(1-\gamma^{5})\,e^{ip\cdot x}\,v_{l}(p,s)\label{Lep_Curr_A0}
\end{equation}
for the leptonic current (here $r$ and $s$ label polarization states),
and
\begin{equation}
\left\langle 0\right|\bar{\psi}_{d}(x)\gamma^{\mu}\gamma^{5}\psi_{u}(x)\left|\pi^{+}(q)\right\rangle \ =\ -\sqrt{2}\,f(q^{2})\,\partial^{\mu}\,e^{-i\,q\cdot x}\ =\ i\,\sqrt{2}\,f(q^{2})\,q^{\mu}\,e^{-i\,q\cdot x}\label{Had_Curr_A0}
\end{equation}
for the hadronic one. Here, the charged lepton and the neutrino are
described as elementary pointlike particles, whereas the pion is treated as
a composite object. Since the pion is a pseudoscalar particle, only the
axial vector piece of the hadronic current contributes to the amplitude. The
internal structure of the pion is taken into account through the form factor
$f(q^{2})$, whose on-shell value is the pion decay constant,
\begin{equation}
f\left(m_{\pi}^{2}\right)\ =\ f_{\pi}\ .
\end{equation}

{}From the above amplitudes, the decay rate for the process $\pi^{+}\rightarrow l^{+}\nu$
is found to be given by the well known expression
\begin{equation}
\Gamma_{\pi^{+}\rightarrow\,l^{+}\nu_{l}}\ =\frac{G_{F}^{2}
\left|V_{\text{ud}}\right|^{2}}{4\pi}\,f_{\pi}^{2}\,m_{\pi}\,
m_{l}^{2}\left(1-\frac{m_{l}^{2}}{m_{\pi}^{2}}\right)^{2}\ .
\label{gammavac}
\end{equation}

\subsection{Background laser field}

We proceed now to describe the features of the external background
field. We consider the case in which it is given by an electromagnetic
plane wave of wave vector $k^{\mu}$, i.e., the electromagnetic
field $A^{\mu}$ has the form
\begin{equation}
A^{\mu}(x)\ =\ e_{1}^{\mu}\,\text{a}_{1}(\zeta)+e_{2}^{\mu}\,\text{a}_{2}(\zeta)\ ,\label{EM_BG}
\end{equation}
where $\zeta=k_{\mu}\,x^{\mu}$, with $k^{2}=0$. Here, $\text{a}_{i}(\zeta)$
are arbitrary functions of $\zeta$, while $e_{i}^{\mu}=(0,\vec{e}_{i})$
are orthogonal polarization vectors (i.e.,\ $\vec{e}_{1}\cdot\vec{e}_{2}=0$)
that satisfy the transversality condition $k\cdot e_{i}=0$. The electromagnetic
strength tensor is given by
\begin{equation}
F^{\mu\nu}\ =\ \partial^{\mu}A^{\nu}-\partial^{\nu}A^{\mu}\ =\ k^{\mu}\,A^{\prime\nu}-k^{\nu}\,A^{\prime\mu}\ ,
\end{equation}
where
\begin{equation}
A^{\prime\mu}\ =\ e_{1}^{\mu}\,\frac{d\text{a}_{1}}{d\zeta}+e_{2}^{\mu}\,\frac{d\text{a}_{2}}{d\zeta}\ .
\end{equation}

As stated above, we focus on the description of the pion decay in
the presence of a background laser field. For a laser pulse of finite
duration, the coefficients $\text{a}_{i}(\zeta)$ in Eq.~(\ref{EM_BG})
would be zero for values of $\zeta$ beyond the interaction region.
Nevertheless, here we will assume that the actual length of the laser
pulse is relatively large, in such a way that it can be approximately
described by an infinitely extended PW. %
Moreover, for the sake of simplicity we will choose a monocromatic,
circularly polarized plane wave (CPPW), taking
\begin{equation}
\text{a}_{1}(\zeta)=a\cos\zeta\ ,\qquad\qquad\text{a}_{2}(\zeta)=-a\sin\zeta\ .
\end{equation}
We can write now $k^{\mu}=\omega\left(1,\vec{n}\right)$, with $\vec{n}^{\,2}=1$,
$\vec{n}\cdot\vec{e}_{i}=0$. Thus, $a$ stands for the wave amplitude,
while $\lambda=2\pi/\omega$ is the corresponding wave length. We
use the vectors $n^{\mu}=(1,\vec{n})/\sqrt{2}$, $\tilde{n}^{\mu}=(1,-\vec{n})/\sqrt{2}$,
$e_{1}^{\mu}$ and $e_{2}^{\mu}$ as an orthonormal basis for space-time.
In addition, we choose the spatial axes in such a way that $\vec{n}$,
$\vec{e}_{1}$ and $\vec{e}_{2}$ point to the $\hat{z}$, $\hat{x}$
and $\hat{y}$ directions, respectively. Then the electromagnetic
field reads
\begin{equation}
A^{\mu}=a\,\left(0,\,\cos\zeta,\,-\sin\zeta,\,0\right)\ .
\label{EM_BG_CPPW1}
\end{equation}
Alternatively, one can introduce the useful notation~\cite{Brown:1964zzb}
\begin{equation}
A^{\mu}(x)\ =\ a\,\text{Re}\left(\ABK^{\mu}\,e^{-i\,\zeta}\right)\ ,
\label{EM_BG_CPPW2}
\end{equation}
with
\begin{equation}
\ABK^{\mu}\ =\ \left(0,\,1,\,-i,\,0\right)\ .\label{A_CPPW}
\end{equation}

It is easily seen that the external electromagnetic field is a function
of $x^{-}=(x^{0}-x^{3})/\sqrt{2}$. As a consequence, the system under
study breaks translational invariance only in the $x^{-}$ direction,
and it is advantageous to describe it using light-front coordinates.
The coordinate $x^{-}$ can be interpreted as the light front time,
while $(x^{+},\vec{x}_{\perp})$, where $x^{+}=(x^{0}+x^{3})/\sqrt{2}$
and $\vec{x}_{\perp}$ lies in the spacial plane perpendicular to
$\vec{n}$ (defined by the vectors $\vec{e}_{i}$), are intended to
be the ``light-front space coordinates''. Concerning the momentum
components, as $x^{-}$ is the ``light-front time coordinate'',
its canonical conjugate $p^{+}=(p^{0}+p^{3})/\sqrt{2}$ can be referred
to as the ``light-front hamiltonian'', and the canonical conjugates
of the ``light-front space coordinates'' $(x^{+},\vec{x}_{\perp})$
will be the ``light-front momentum coordinates'' $(p^{-},\vec{p}_{\perp})$,
where $p^{-}=(p^{0}-p^{3})/\sqrt{2}$. For a particle of mass $m$
on its mass shell we have $p^{+}=(m^{2}+\vec{p}_{\perp}^{\;2})/(2p^{-})$.

To fully characterize the charged pion decay in the presence of this
background field we also need to specify the pion four-momentum. Several
dimensionless, scalar and gauge invariant parameters have been defined in
the literature for the description of this system. A discussion about the
parameters of interest and the corresponding orders of magnitude can be
found in Refs.~\cite{Fedotov:2022ely,Seipt:2017ckc,Mitter:1974yg}. We follow
Ref.~\cite{Fedotov:2022ely} for a manifestly gauge-invariant definition of
these parameters, changing the electron mass used in
Ref.~\cite{Fedotov:2022ely} for the pion mass. Thus, we define the classical
non linearity parameter $\xi_{\pi}$ as
\begin{equation}
\xi_{\pi}\ =\ \frac{Q_{\pi}}{m_{\pi}\ k\cdot q}\ \left\langle -\,q_{\mu}F^{\mu\nu}q^{\alpha}F_{\alpha\nu}\right\rangle ^{1/2}\ ,\label{chi}
\end{equation}
where $q^{\mu}$ is the momentum of the decaying pion, $Q_{\pi}$
is the pion electric charge 
and the angle brackets indicate a phase cycle average. On the other
hand, the relation between the energy of the probe particle and the
plane wave background is parameterized by the linear quantum parameter
\begin{equation}
\eta_{\pi}\ =\ \frac{k\cdot q}{m_{\pi}^{2}}\ .\label{eta}
\end{equation}
In our case, the background field can be characterized by two parameters:
its intensity, $a$, and the frequency of the laser field in the lab
frame, $\omega$ (or, equivalently, the $+$-component of the wave vector,
$k^{+}$). These parameters can be written in terms of the dimensionless
quantities given in Eqs.~(\ref{chi})
and (\ref{eta}) as %

\begin{equation}
Q_{\pi}\,a=\xi_{\pi}\,m_{\pi}\ ,\qquad\qquad\sqrt{2}\,\omega\,q^{-}=\eta_{\pi}\,m_{\pi}^{2}\ .\label{Def_=00003D00005Cchi_=00003D00005Ceta}
\end{equation}
In addition, as stated above, we consider the case in which the pion
momentum points in the direction of the wave vector, i.e.\ we take
$\vec{q}_{\perp}=0$.

\newcounter{eraII}
\renewcommand{\thesection}{\arabic{eraII}}
\renewcommand{\theequation}{\arabic{eraII}.\arabic{equation}}
\setcounter{eraII}{3} \setcounter{equation}{0}

\section{Leptonic and hadronic matrix elements in the presence of the external
field }

\subsection{Pion-to-vacuum hadronic amplitude}

\label{hadamp}

In the presence of the external electromagnetic field the decay process
$\pi^{+}\rightarrow l^{+}\nu_{l}$ can still be written in terms of a Fermi
Lagrangian density as the one in Eq.~(\ref{FermiLagran}). However, charged
particle fields have to be modified, and a parametrization of the
pion-to-vacuum hadronic amplitude in terms of several form factors is
expected.

The hadronic current $J_{h}^{\mu}(x)$ in Eq.~(\ref{currents}) includes
vector and axial vector pieces. In vacuum, the vector piece
does not contribute to the pion-to-vacuum amplitude, while the contribution
of the axial vector piece is given by Eq.~(\ref{Had_Curr_A0}). Now,
in the presence of an electromagnetic background field $A^{\mu}$,
new tensors are available and can be used to obtain the desired Lorentz
structure. In the most general situation, due to gauge symmetry, the
derivative $\partial^{\mu}$ has to be replaced by the covariant derivative,
\begin{equation}
D^{\mu}\ =\ \partial^{\mu}+i\,Q_{\pi}A^{\mu}\ .\label{dmu}
\end{equation}
In addition, one has other two gauge invariant tensors, namely, the
electromagnetic strength tensor and its dual,
\begin{equation}
F^{\mu\nu}=\partial^{\mu}A^{\nu}-\partial^{\nu}A^{\mu}\ ,\qquad\qquad\widetilde{F}^{\mu\nu}=\frac{1}{2}\,\epsilon^{\mu\nu\rho\sigma}\,F_{\rho\sigma}\ .\label{efes}
\end{equation}
The vector and axial vector contributions to the pion-to-vacuum amplitude
have to be built from contractions between these objects. From the
strength tensor and its dual one can obtain the relativistic invariant
quantities
\begin{align}
\mathcal{S} & \equiv\frac{1}{4}\,F_{\mu\nu}F^{\mu\nu}=-\frac{1}{4}\,\widetilde{F}_{\mu\nu}\widetilde{F}^{\mu\nu}=\frac{1}{2}\,\left(E^{2}-B^{2}\right)\ ,\nonumber \\
\mathcal{P} & \equiv\frac{1}{4}\,F_{\mu\nu}\widetilde{F}^{\mu\nu}=\vec{E}\cdot\vec{B}\ ,
\end{align}
where $\vec{E}$ and $\vec{B}$ are the external electric and magnetic
fields. Other useful relations are the following: taking two tensors
and contracting one Lorentz index one gets
\begin{align}
F^{\mu\alpha}F_{\alpha\nu} & \ =\ \widetilde{F}^{\mu\alpha}\widetilde{F}_{\alpha\nu}-2\,\mathcal{S}\,\delta_{\nu}^{\mu}\ ,\nonumber \\
\widetilde{F}^{\mu\alpha}F_{\alpha\nu} & \ =\-\,\mathcal{P}\,\delta_{\nu}^{\mu}\ ,
\end{align}
while covariant derivatives satisfy the commutation relations
\begin{align}
[D_{\mu}\,,\,D_{\nu}] & \ =\ i\,Q_{\pi}\,F^{\mu\nu}\ ,\nonumber \\[2mm]{}
[D_{\alpha}\,,\,F^{\mu\nu}] & \ =\ \partial_{\alpha}F^{\mu\nu}\ .
\end{align}

Let us consider, in particular, an external electromagnetic PW. It
is easy to see that in this case one has ${\cal S}={\cal P}=0$. Taking
into account the previous relations, from the available tensors one
can set up just three independent vector operators,
\begin{equation}
D^{\mu}\ ,\quad Q_{\pi}\,F^{\mu\nu}D_{\nu}\ ,\quad Q_{\pi}^{2}\,F^{\mu\alpha}F_{\alpha\nu}\,D^{\nu}\ ,
\end{equation}
and one axial vector operator,
\begin{equation}
Q_{\pi}\,\widetilde{F}^{\mu\nu}D_{\nu}\ .
\end{equation}
In addition, in the case of an external electromagnetic PW, another
available vector-like physical quantity is the wave vector $k^\mu$.

On the other hand, particle fields get modified in the
presence of the external field. The form of the wavefunctions $\phi_{q}^{\pm}(x)$
for charged pions is discussed in App.~\ref{sec:Particle-fields}.
In the case of an electromagnetic PW, they can be written as
\begin{equation}
\phi_{q}^{\pm}(x)\ =\ e^{\mp iq\cdot x}\,\varphi_{q}^{\pm}(\zeta)\ ,
\end{equation}
where $q^\mu = (E_q,\vec q\,)$, with $E_q = \sqrt{\vec q^{\,\,2}+m_\pi^2}$,
and the functions $\varphi_{q}^{\pm}(\zeta)$ are given by Eqs.~(\ref{fi+})
and (\ref{fi-}).

Thus, the vector and axial vector pieces of the pion-to-vacuum amplitude
can be written as
\begin{align}
\langle0|\bar{\psi}_{d}(x)\gamma^{\mu}\psi_{u}(x)|\pi^{+}(q)\rangle & \ =\
\sqrt{2}\,f_{V}\,Q_{\pi}\,\widetilde{F}^{\mu\nu}D_{\nu}\,\phi_{q}^{+}(x)\ ,\label{vector}\\
\langle0|\bar{\psi}_{d}(x)\gamma^{\mu}\gamma^{5}\psi_{u}(x)|\pi^{+}(q)\rangle & \ =
\ -\sqrt{2}\,\left[f_{A1}\,D^{\mu}+\, i f_{A2}\,Q_{\pi}\,F^{\mu\nu}D_{\nu}+
f_{A3}\,Q_{\pi}^{2}\,F^{\mu\alpha}F_{\alpha\nu}\,D^{\nu}\right]\phi_{q}^{+}(x)\ ,
\label{axial-vector}
\end{align}
where $Q_{\pi}$ is equal to the positron charge $e$. In principle, in
Eq.~(\ref{axial-vector}) one could include a further term proportional to
$k^\mu$; however, as shown below, the corresponding form factor can be
absorbed through a redefinition of $f_{A3}$.

The form factors $f_{V}$ and $f_{Ai}$, $i=1,2,3$, are scalar functions of
the Lorentz invariant differential operators that can be built using the
gauge invariant tensors in Eqs.~(\ref{dmu}) and (\ref{efes}) together
with the wave vector $k^\mu$. In fact, it can be seen that only  three
independent nontrivial scalar operators can be found. They can be chosen to
be e.g.
\begin{equation}
D^{\mu}D_{\mu}\ ,\qquad\qquad Q_{\pi}^{2}\,F^{\mu\nu}F_{\alpha\nu}\,D^{\alpha}D_{\mu}
\ , \qquad\qquad k^{\mu}D_{\mu}\ .
\end{equation}
Taking into account the relation
\begin{equation}
D_{\mu}\phi_{q}^{+}(x)\ =\;\left[-\,i\,q_{\mu}-\,\frac{i}{k\cdot q}\,\Big(Q_{\pi}\,A(\zeta)\cdot q-\frac{1}{2}\,Q_{\pi}^{2}\,A(\zeta)^{2}\Big)\,k_{\mu}+iQ_{\pi}\,A_{\mu}(\zeta)\right]\phi_{q}^{+}(x)\label{dmuphi}
\end{equation}
(see App.~\ref{sec:Particle-fields}), it is seen that the action
of these scalar operators on the pion wave function gives
\begin{align}
D^{\mu}D_{\mu}\,\phi_{q}^{+}(x) \ & = \ -q^{2}\,\phi_{q}^{+}(x)\ ,\nonumber \\[2mm]
Q_{\pi}^{2}\,F^{\mu\nu}F_{\alpha\nu}\,D^{\alpha}D_{\mu}\,\phi_{q}^{+}(x) \ & =
 \ \ - Q_{\pi}^{2}\,(k\cdot q)^{2}{A'}^{2}\,\phi_{q}^{+}(x)\ , \nonumber \\[2mm]
 k^{\mu}D_{\mu}\,\phi_{q}^{+}(x) \ &  \ = \ -i\, k\cdot q\,\phi_{q}^{+}(x)\ .
\label{scalars}
\end{align}
Hence, the form factors will be in general functions of the Lorentz
invariant quantities $q^{2}$,  $k\cdot q$ and $Q_{\pi}^{2}\,(k\cdot
q)^{2}{A'}^{2}$.

From Eqs.~(\ref{vector}), (\ref{axial-vector}) and (\ref{dmuphi})
the pion-to-vacuum matrix element of the hadronic $V-A$ current is
given by
\begin{align}
\left\langle 0\right|J_{h}^{\mu}(x)\left|\pi^{+}(q)\right\rangle  & \,=\,\left\langle 0\right|\bar{\psi}_{d}(x)\gamma^{\mu}(1-\gamma^{5})\psi_{u}(x)\left|\pi^{+}(q)\right\rangle \nonumber \\[2mm]
 & \,=\,\sqrt{2}\,e^{-iq\cdot x}\,\varphi_{q}^{+}(\zeta)\;\mathbb{F}^{\mu}\ ,\label{hadcur}
\end{align}
where $\varphi_{q}^{+}(\zeta)$ is given by Eq.~(\ref{fi+}), and
the vector $\mathbb{F}^{\mu}$ is defined as
\begin{align}
\mathbb{F}^{\mu}\ =\  & -i\left(f_{V}\,g_{V}^{\mu}+f_{A1}\,g_{A1}^{\mu}+ if_{A2}\,g_{A2}^{\mu}+
f_{A3}\,g_{A3}^{\mu}\right)\ ,
\label{3.30-1}
\end{align}
with
\begin{align}
g_{V}^{\mu}= & \ Q_{\pi}\,\epsilon^{\mu\nu\rho\sigma}\,k_{\rho}\,A_{\sigma}^{\prime}\left(q_{\nu}-Q_{\pi}\,A_{\nu}\right)\ ,\nonumber \\[2mm]
g_{A1}^{\mu}= & \ q^{\mu}+\frac{Q_{\pi}}{k\cdot q}\ A\cdot\left(q-Q_{\pi}\,\frac{A}{2}\right)\,k^{\mu}-Q_{\pi}\,A^{\mu}\ ,\nonumber \\[2mm]
g_{A2}^{\mu}= & \ Q_{\pi}\,A^{\prime}\cdot\left(q-Q_{\pi}\,A\right)\,k^{\mu}-Q_{\pi}\,k\cdot q\ A^{\prime\mu}\ ,\nonumber \\[2mm]
g_{A3}^{\mu}= & \ -\,Q_{\pi}^{2}\ A^{\prime}\cdot A^{\prime}\ k\cdot q\,k^{\mu}\ .\label{gVA123_Gen}
\end{align}
In this way, as pointed out in Ref.~\cite{Coppola:2018ygv}, it is
seen that the presence of an electromagnetic background field enables
the existence of new form factors for pion decays. In particular,
in our case we have a vector form factor $f_{V}$ and three axial
form factors $f_{Ai}$. In the $A^{\mu}\rightarrow0$ limit the coefficients
accompanying the new form factors $f_{V}$, $f_{A2}$ and $f_{A3}$
vanish, whereas $g_{A1}^{\mu}\to q^{\mu}$ and $f_{A1}\rightarrow f_{\pi}$.

In the particular case of a CPPW, it is seen that Eqs.~(\ref{gVA123_Gen})
get further simplified, since one has $A\cdot A'=0$, $A^{2}={A^{\prime}}^{2}=-a^{2}$.
Moreover, if we consider the case in which $\vec{q}_{\perp}=0$, we
also have $q\cdot A=q\cdot A'=0$. Eqs.~(\ref{gVA123_Gen}) then
reduce to
\begin{align}
g_{V}^{\mu}= & \ Q_{\pi}\,\epsilon^{\mu\nu\rho\sigma}\,k_{\rho}\,A_{\sigma}^{\prime}(q_{\nu}-Q_{\pi}\,A_{\nu})\,\nonumber \\[2mm]
g_{A1}^{\mu}= & \ q^{\mu}-Q_{\pi}\,A^{\mu}+\,\frac{\xi_{\pi}^{2}}{2\,\eta_{\pi}}\,k^{\mu}\ ,\nonumber \\[2mm]
g_{A2}^{\mu}= & \ -\eta_{\pi}\,m_{\pi}^{2}\,Q_{\pi}\,A^{\prime\mu}\ ,\nonumber \\[2mm]
g_{A3}^{\mu}= & \ \xi_{\pi}^{2}\,\eta_{\pi}\,m_{\pi}^{4}\,k^{\mu}\ .\label{gVA123_CPPW}
\end{align}
Assuming that the pion is on shell, the invariants in Eqs.~(\ref{scalars})
are given by
\begin{equation}
q^{2}=m_{\pi}^{2}\ ,\qquad\qquad
 k\cdot q = m_\pi^2\,\eta_\pi\ ,\qquad\qquad
-Q_{\pi}^{2}\,(k\cdot q)^{2}{A'}^{2}=(\xi_{\pi}\,\eta_{\pi}\,m_{\pi}^{3})^{2}\ .\label{invariants}
\end{equation}

\subsection{Leptonic amplitude}

In the case of the charged leptons, the form of the corresponding
wavefunctions in the presence of an electromagnetic PW background
is given in App.~\ref{app_a_2}, where some definitions are introduced.
Taking this into account, we get
\begin{equation}
\left\langle \bar{l}({p},s)\nu_{l}({t},r)\right|\bar{\Psi}_{\nu_{l}}(x)\gamma^{\mu}(1-\gamma^{5})\Psi_{l}(x)\left|0\right\rangle \ =\ e^{it\cdot x}\,\bar{u}_{\nu}({t},r)\,\gamma^{\mu}(1-\gamma^{5})\,\boldsymbol{E}_{p}^{c}(x)\,v_{l}({p},s)\ ,\label{lepmatrix}
\end{equation}
where
\begin{equation}
\boldsymbol{E}_{p}^{c}(x)\ =\ e^{ip\cdot x}\,\varphi_{p}^{-}(\zeta)\,\left(1-\frac{Q_{l}}{2\,k\cdot p}\,\cancel{k}\,\cancel{A}(\zeta)\right)\ ,
\end{equation}
with $\varphi_{p}^{-}(\zeta)$ given by Eq.~(\ref{fi-}). Here, $\boldsymbol{E}_{p}^{c}(x)$
is the Ritus matrix~\cite{Seipt:2017ckc,Ritus:1978cj} associated
with the creation of the $l^{+}$ particle in the final state, and (with
our convention for particles and antiparticles) $Q_{l}=-e$. In fact,
for convenience, we have removed in Eq.~(\ref{lepmatrix}) a prefactor
$(2\pi)^{-3}\,(2E_{p}\,2E_{t})^{-1/2}$, which will be taken into
account later through the phase space factors in the decay width.

\newcounter{eraIII}
\renewcommand{\thesection}{\arabic{eraIII}}
\renewcommand{\theequation}{\arabic{eraIII}.\arabic{equation}}
\setcounter{eraIII}{4} \setcounter{equation}{0}

\section{$\pi^{+}\rightarrow l^{+}\nu_{l}$ decay amplitude and width}

\subsection{$\pi^{+}\rightarrow l^{+}\nu_{l}$ decay amplitude}

The transition amplitude for the process $\pi^{+}(q)\rightarrow l^{+}(p)\,\nu_{l}(t)$
is given by
\begin{equation}
S_{\pi^{+}\rightarrow\,l^{+}\nu_{l}}\ =\ -i\,\frac{G_{F}\,V_{\text{ud}}}{\sqrt{2}}\,\int d^{4}x\,\langle0|J_{h}^{\mu}(x)|\pi^{+}({q})\rangle\,\langle\bar{l}({p},s)\nu_{l}({t},r)|J_{l\mu}^{\dagger}(x)|0\rangle\ .
\end{equation}
As discussed in the previous subsections, in the case of an electromagnetic
CPPW background the pion-to-vacuum matrix element of the hadronic
current can be written as in Eq.~(\ref{hadcur}), while the leptonic
matrix element is given by Eq.~(\ref{lepmatrix}). One has in this
way
\begin{equation}
S_{\pi^{+}\rightarrow l^{+}\nu_{l}}\ =\ -i\,\int d^{4}x\,e^{i\left(p+t-q\right)\cdot x}\,I_{pq}(\zeta)\;\mathbb{M}_{({p},s)({t},r)}(\zeta)\ ,\label{eqspiplus}
\end{equation}
where we have defined
\begin{equation}
I_{pq}(\zeta)\ =\ \varphi_{p}^{(-)}(\zeta)\;\varphi_{q}^{(+)}(\zeta)
\end{equation}
and
\begin{equation}
\mathbb{M}_{({p},s)({t},r)}(\zeta)\ =\ G_{F}\,V_{\text{ud}}\,\bar{u}_{\nu}({t},r)\,\cancel{\mathbb{F}}\,(1-\gamma^{5})\,\left(1-\frac{Q_{l}}{2\,k\cdot p}\,\cancel{k}\,\cancel{A}(\zeta)\right)\,v_{l}({p},s)\ ,
\label{FermiContr}
\end{equation}
with $\varphi_{p}^{-}(\zeta)$, $\varphi_{q}^{+}(\zeta)$ given by
Eqs.~(\ref{fi+}) and (\ref{fi-}), and $\mathbb{F}^{\mu}$ given
by Eqs.~(\ref{3.30-1}) and (\ref{gVA123_CPPW}).

We follow now the procedure of evaluation carried out in Ref.~\cite{Brown:1964zzb}.
In the case of a CPPW, and taking $\vec{q}_{\perp}=0$, we have
\begin{equation}
\varphi_{q}^{(+)}(\zeta)=\exp\left[-\frac{i}{k\cdot q}
\int_{-\infty}^{\zeta}\,d\zeta'\,\frac{1}{2}\,e^{2}\,a^{2}\right]\ ,\qquad
\varphi_{p}^{(-)}(\zeta)=\exp\left[\frac{i}{k\cdot p}
\int_{-\infty}^{\zeta}\,d\zeta'\,\left(e\,A(\zeta')\cdot p+\frac{1}{2}\,e^{2}\,a^{2}\right)\right]\ .
\end{equation}
Thus, using the expression in Eq.~(\ref{EM_BG_CPPW2}) for the
electromagnetic field, and assuming that the latter is switched on/off
adiabatically, we can perform the integrals obtaining
\begin{align}
I_{pq}(\zeta)\  =  \exp & \left[-\frac{e\,a}{2}\left(\frac{\ABK\cdot p}{k\cdot p}\,
e^{-i\,\zeta}-\frac{\ABK^{*}\cdot p}{k\cdot p}\,e^{i\,\zeta}\right)+
i\,\frac{e^{2}\,a^{2}}{2}\,\left(\frac{1}{k\cdot p}-\frac{1}{k\cdot q}\right)\,\zeta\right]\ ,
\end{align}
which can also be written as
\begin{align}
I_{pq}(\zeta)\ =\ \exp & \left[-i\,d_{1}\sin(\zeta-\alpha)+i\,d_{2}\,\zeta\right]\ ,
\end{align}
with the definitions
\begin{eqnarray}
d_{1}=e\,a\,\frac{|\ABK\cdot p|}{k\cdot p}\ ,
\qquad\quad
\alpha={\rm arg}(-\ABK\cdot p)\ ,\qquad\quad d_{2}=\frac{e^{2}\,a^{2}}{2}\,\left(\frac{1}{k\cdot p}-\frac{1}{k\cdot q}\right)\ .
\label{d1_d2_0}
\end{eqnarray}
Using polar coordinates in the perpendicular plane, i.e.,
\begin{equation}
p^{1}=p_{\perp}\cos\theta\ ,\qquad\qquad\qquad p^{2}=p_{\perp}\sin\theta\ ,
\end{equation}
and taking into account the parameters introduced in Eqs.~(\ref{chi})
and (\ref{eta}), we have
\begin{equation}
d_{1}=\frac{\xi_{\pi}}{\eta_{\pi}}\,\frac{k\cdot q}{k\cdot p}\,\frac{p_{\perp}}{m_{\pi}}\ ,\qquad\quad\alpha=-\theta\ ,\qquad\quad d_{2}=\frac{\xi_{\pi}^{2}}{2\eta_{\pi}}\,\left(\frac{k\cdot q}{k\cdot p}-1\right)\ .\label{d1_d2_1}
\end{equation}

Now, we can use the generating function of the Bessel functions
\begin{equation}
e^{i\,z\sin\beta}=\sum_{\ell=-\infty}^{+\infty}\,e^{i\,\ell\beta}\,J_{\ell}(z)
\end{equation}
to express $I_{pq}$ as
\begin{equation}
I_{pq}(\zeta)\ =\ e^{id_{2}\,\zeta}\sum_{\ell=-\infty}^{+\infty}\,e^{-i\,\ell\,(\zeta-\alpha)}\,J_{\ell}(d_{1})\ .\label{3.50}
\end{equation}
Replacing in Eq.~(\ref{eqspiplus}), we obtain
\begin{align}
S_{\pi^{+}\rightarrow l^{+}\nu_{l}}\ =\ -i\int d^{4}x & \,e^{i\left(p+t-q\right)\cdot x}\,e^{i\,d_{2}\,\zeta}\sum_{\ell=-\infty}^{+\infty}\,e^{-i\,\ell\,(\zeta+\theta)}\,J_{\ell}(d_{1})\ \mathbb{M}_{({p},s)({t},r)}(\zeta)\ .\label{3.51}
\end{align}
Next, we evaluate the fermionic contribution
$\mathbb{M}_{({p},s)({t},r)}(\zeta)$, given in Eq.~(\ref{FermiContr}), where
$Q_{l}=- e$ and $\mathbb{F}^{\mu}$ is given by Eqs.~(\ref{3.30-1}) and
(\ref{gVA123_CPPW}). After some calculation we arrive at the expression
\begin{align}
\mathbb{M}_{({p},s)({t},r)}(\zeta) & =\ G_{F}\,V_{\text{ud}}\,
\left(\mathbb{M}^{\left(0\right)}+\mathbb{M}^{\left(+1\right)}\,
e^{-i\,\zeta}+\mathbb{M}^{\left(-1\right)}\,e^{i\,\zeta}\right)\ ,\label{3.70}
\end{align}
where $\mathbb{M}^{\left(i\right)}$ are $\zeta$-independent spinor matrix elements given by
\begin{align}
\mathbb{M}^{\left(0\right)}= & \ i\,\bar{u}_{\nu}({t},r)\left\{ -f_{A1}\,\cancel{q}+\frac{e^{2}\,a^{2}}{2}
\left[\left(\frac{1}{k\cdot p}-\frac{1}{k\cdot q}\right)\,f_{A1}+\frac{k\cdot q}{k\cdot p}\,f_{VA-}-
2f_{VA3}\right]\,\cancel{k}\right\} \,(1-\gamma^{5})\,v_{l}({p},s)\ ,\nonumber \\[2mm]
\mathbb{M}^{\left(+1\right)}= & \ i\,\frac{e\,a}{2}\,\left[f_{A1}\frac{k\cdot\left(p-q\right)}{k\cdot p}+
k\cdot q\ f_{VA+}\right]\,\bar{u}_{\nu}({t},r)\,\cancel{\ABK}\,(1-\gamma^{5})\,v_{l}({p},s)\ ,\nonumber \\[2mm]
\mathbb{M}^{\left(-1\right)}= & \ i\,\frac{e\,a}{2}\,\left[f_{A1}+k\cdot q\
f_{VA-}\right]\,\bar{u}_{\nu}({t},r)\,\cancel{\ABK}^{\,*}\,(1-\gamma^{5})\,v_{l}({p},s)\ ,
\end{align}
with
\begin{align}
f_{VA+} & =f_{V}+f_{A2}\ ,\nonumber \\
f_{VA-} & =f_{V}-f_{A2}\ ,\nonumber \\
f_{VA3} & =f_{V}+k\cdot q\ f_{A3}\ .
\label{deff}
\end{align}
Finally, changing $\ell\to\ell\pm1$ conveniently in the sums, we
see that the $\pi^{+}\rightarrow l^{+}\nu_{l}$ transition amplitude
can be written as
\begin{align}
S_{\pi^{+}\rightarrow l^{+}\nu_{l}}= & \sum_{\ell=-\infty}^{+\infty}\,\int d^{4}x\,e^{i\left(p+t-q\right)\cdot x}\,e^{i\,d_{2}\,\zeta}\,e^{-i\,\ell\,\zeta}\,\mathbb{M}_{\ell}\nonumber \\
= & \sum_{\ell=-\infty}^{+\infty}\,\left(2\pi\right)^{4}\,\delta^{(4)}(p+t-q-(\ell-d_{2})\,k)\ \mathbb{M}_{\ell}\ ,\label{scondelta}
\end{align}
where
\begin{equation}
\mathbb{M}_{\ell}\ =\ -i\,G_{F}\,V_{\text{ud}}\,e^{-i\,\ell\theta}\,\left[J_{\ell}(d_{1})\,\mathbb{M}^{\left(0\right)}+e^{i\theta}J_{\ell-1}(d_{1})\,\mathbb{M}^{\left(+1\right)}+e^{-i\theta}J_{\ell+1}(d_{1})\,\mathbb{M}^{\left(-1\right)}\right]\ .\label{M_l_Gorda}
\end{equation}
In this way, it is seen that the total transition amplitude can be
understood as the sum of amplitudes that describe the disintegration
of the pion into the charged lepton and its neutrino together with a number
$\ell$ of background photons. The energy-momentum conservation gets
expressed by the condition
\begin{equation}
p+t-q-\left(\ell-d_{2}\right)\,k\ =\ 0\ .\label{MomentumConserv}
\end{equation}

\subsection{$\pi^{+}\rightarrow l^{+}\nu_{l}$ decay width}

The decay width $\Gamma(\pi^{+}\rightarrow l^{+}\nu_{l})$ can be
calculated from the transition amplitude $S_{\pi^{+}\rightarrow\,l^{+}\nu_{l}}$
obtained in the previous subsection. Taking into account the $\ell$
dependence of the momentum conservation delta in Eq.~(\ref{scondelta}),
it is seen that only terms of the form $\left|\mathbb{M}_{\ell}\right|^{2}$
can appear in the differential decay width. Thus, we have
\begin{equation}
d\Gamma(\pi^{+}\rightarrow l^{+}\nu_{l})\ =\ \sum_{\ell=-\infty}^{\infty}\,(2\pi)^{4}\,
\delta^{(4)}(p+t-q-\left(\ell-d_{2}\right)\,k)\,\frac{1}{2E_{q}}\,\frac{d^{3}t}{(2\pi)^{3}\,2E_{t}}\,
\frac{d^{3}p}{(2\pi)^{3}\,2E_{p}}\;\overline{\left|\mathbb{M}_{\ell}\right|^{2}}\ ,\label{sumal}
\end{equation}
where $\overline{\left|\mathbb{M}_{\ell}\right|^{2}}=\sum_{r,s}|\mathbb{M}_{\ell}|^{2}$.
It is convenient to express the phase space factor in terms of light
front coordinates, defining
\begin{equation}
d\Phi_{\ell}\ =\ (2\pi)^{4}\,\delta^{(4)}(p+t-q-(\ell-d_{2})\,k)\,\frac{dt^{-}\,d^{2}t_{\perp}}{(2\pi)^{3}\,2\,t^{-}}\,\frac{dp^{-}\,d^{2}p_{\perp}}{(2\pi)^{3}\,2\,p^{-}}\ ,\label{2.177}
\end{equation}
where $d^{2}p_{\perp}=dp^{1}\,dp^{2}$, $d^{2}t_{\perp}=dt^{1}\,dt^{2}$.
One has then
\begin{equation}
d\Gamma(\pi^{+}\rightarrow l^{+}\nu_{l})\ =\ \frac{1}{2E_{q}}\,\sum_{\ell=-\infty}^{\infty}\;d\Phi_{\ell}\,\overline{\left|\mathbb{M}_{\ell}\right|^{2}}\ .\label{3.79}
\end{equation}
It is also useful to introduce a variable $x$ that denotes the fraction
of the light-front momentum of the pion ($q^{-}$) that is carried
away by the charged lepton, i.e.,
\begin{equation}
x\ =\ \frac{p^{-}}{q^{-}}\ ,\label{3.58}
\end{equation}
with $0\leq x\leq1$. After some algebra we obtain
\begin{align}
d\Phi_{\ell}\ =\  & \frac{1}{4(2\pi)^{2}}\;dx\,d\theta\;dp_{\perp}^{2}\,\delta(p_{\perp}^{2}-(1-x)\,(x\,m_{\pi}^{2}\,(1+2\,\ell\,\eta_{\pi})-(1-x)\,m_{\pi}^{2}\,\xi_{\pi}^{2}-m_{l}^{2}))\nonumber \\
 & \times dt^{-}\,\delta(t^{-}+p^{-}-q^{-})\,d^{2}t_{\perp}\,\delta^{(2)}(\vec{t}_{\perp}+\vec{p}_{\perp})\ .\label{3.57}
\end{align}
In terms of $\vec{p}_{\perp}$, $x$ and $q^{-}$, particle momenta
can be written as
\begin{equation}
p^{+}=\frac{m_{l}^{2}+p_{\perp}^{2}}{2\,x\,q^{-}}\ ,
\qquad t^{-}=\left(1-x\right)q^{-}\ ,
\qquad\vec{t}_{\perp}=-\vec{p}_{\perp}\ ,
\qquad t^{+}=\frac{p_{\perp}^{2}}{2\,(1-x)\,q^{-}}\ ,
\qquad q^{+}=\frac{m_{\pi}^{2}}{2\,q^{-}}\ .\label{3.59}
\end{equation}
The conditions $p_{\perp}^{2}\geq0$ and $0\leq x\leq 1$ imply a lower limit
for the sum over the integer index $\ell$, as it will be seen below.

Using the momentum conservation, Eq.~(\ref{MomentumConserv}), together
with the relations $(\cancel{p}+m_{l})\,v_{l}({p},s)=0$ and $\bar{u}_{\nu_{l}}({t},r)\,\cancel{t}=0$,
we obtain
\begin{align}
\mathbb{M}_{\ell}\ =\  & G_{F}\,V_{\text{ud}}\,e^{-i\,\ell\theta}\,\bar{u}_{\nu}({t},r)\,(1+\gamma^{5})\left(m_{l}\,J_{\ell}\left(d_{1}\right)\,f_{A}^{\left(1\right)}+\cancel{Q}\right)v_{l}({p},s)\ ,
\end{align}
where we have introduced the vector
\begin{align}
Q^{\mu}\ =\  & J_{\ell}(d_{1})\left(\ell\,f_{A1}+
\frac{\xi_{\pi}^{2}\,m_{\pi}^{2}}{2x}\, f_{VA-}-\xi_{\pi}^{2}\,m_{\pi}^{2}\,f_{VA3}\right)k^{\mu}\nonumber \\
 & +e^{i\theta}J_{\ell-1}(d_{1})\,\frac{\xi_{\pi}\,m_{\pi}}{2}\,\left[\left(1-\frac{1}{x}\right)f_{A1}+
 \eta_{\pi}\,m_{\pi}^{2}\, f_{VA+}\right]\,\ABK^{\mu}\nonumber \\
 & +e^{-i\theta}J_{\ell+1}(d_{1})\,\frac{\xi_{\pi}\,m_{\pi}}{2}\,\left(f_{A1}+
 \eta_{\pi}\,m_{\pi}^{2}\, f_{VA-}\right)\ABK^{*\mu}\ .
\label{3.109}
\end{align}
This leads to
\begin{align}
\overline{\left|\mathbb{M}_{\ell}\right|^{2}}\ =\  &
 G_{F}^{2}\,|V_{\text{ud}}|^{2}\;\text{tr}\left[\cancel{t}\,\left(1+\gamma_{5}\right)\,\left(m_{l}\,J_{\ell}(d_{1})\,f_{A1}+\cancel{Q}\right)\left(\cancel{p}-m_{l}\right)\,\left(m_{l}\,J_{\ell}(d_{1})\,f_{A1}^{\,\ast}+\cancel{Q}^{\,\ast}\right)\left(1-\gamma_{5}\right)\right]\ .
\end{align}
After a long and straightforward calculation we arrive at the expression
\begin{equation}
\overline{\left|\mathbb{M}_{\ell}\right|^{2}}=8\,G_{F}^{2}\,
\left|V_{\text{ud}}\right|^{2}\,\sum_{\Sigma,\Sigma'}
\, \epsilon_{\Sigma,\Sigma'}\;\mathbb{N}_{\ell}^{(\Sigma,\Sigma')}\,
\text{Re}\left[f_{\Sigma}\,f_{\Sigma'}^{\,\ast}\right]\ ,
\label{defepsilon}
\end{equation}
where the indices $\Sigma$ and $\Sigma'$ run over $A1$, $VA+$, $VA-$ and
$VA3$.  A factor $\epsilon_{\Sigma,\Sigma'}$, equal to 1 for $\Sigma
=\Sigma'$ and to 1/2 for $\Sigma \neq \Sigma'$, has been included so as to
avoid double counting of cross terms. The functions
$\mathbb{N}_{\ell}^{(\Sigma,\Sigma')}(x,p_{\perp})$, which satisfy
$\mathbb{N}_{\ell}^{(\Sigma,\Sigma')}(x,p_{\perp})=
\mathbb{N}_{\ell}^{(\Sigma',\Sigma)}(x,p_{\perp})$, are given by
\begin{align}
\mathbb{N}_{\ell}^{\left(A1,A1\right)}\ =\  & \frac{m_{l}^{2}}{2\,x}\,\left\{ m_{l}^{2}\,(1-x)\,J_{\ell}(d_{1})^{2}+\frac{1}{1-x}\,\Big[p_{\perp}J_{\ell}(d_{1})-\xi_{\pi}\,m_{\pi}\,(1-x)\,J_{\ell+1}(d_{1})\Big]^{2}\right\} \ ,\label{N_A1_A1}\\
\mathbb{N}_{\ell}^{\left(A1,VA+\right)}\ =\  & -\,m_{l}^{2}\,\eta_{\pi}\,\xi_{\pi}\,m_{\pi}^{3}\,p_{\perp}\,J_{\ell}(d_{1})\,J_{\ell-1}(d_{1})\ ,\\
\mathbb{N}_{\ell}^{\left(A1,VA-\right)}\ =\  & -\,m_{l}^{2}\,\eta_{\pi}\,\xi_{\pi}\,m_{\pi}^{3}\,\left[p_{\perp}\,J_{\ell+1}(d_{1})\,J_{\ell}(d_{1})-\xi_{\pi}\,m_{\pi}\,\frac{(1-x)}{x}\,\left(J_{\ell+1}(d_{1})^{2}-J_{\ell}(d_{1})^{2}\right)\right]\ ,\\
\mathbb{N}_{\ell}^{\left(A1,VA3\right)}\ =\  & 2\,m_{l}^{2}\,\eta_{\pi}\,\xi_{\pi}^{2}\,m_{\pi}^{4}\,(1-x)\,J_{\ell}(d_{1})^{2}\;,\\
\mathbb{N}_{\ell}^{\left(VA+,VA+\right)}\ =\  & \frac{1}{2}\,\eta_{\pi}^{2}\,\xi_{\pi}^{2}\,m_{\pi}^{6}\,\frac{x}{1-x}\,p_{\perp}^{2}\,J_{\ell-1}(d_{1})^{2}\;,\\
\mathbb{N}_{\ell}^{\left(VA-,VA-\right)}\ =\  & \frac{1}{2}\,\eta_{\pi}^{2}\,\xi_{\pi}^{2}\,m_{\pi}^{6}\,\frac{(1-x)}{x}\,\left[m_{l}^{2}\,J_{\ell+1}(d_{1})^{2}+\Big(p_{\perp}\,J_{\ell+1}(d_{1})-\xi_{\pi}\,m_{\pi}\,J_{\ell}(d_{1})\Big)^{2}\right]\ ,\\
\mathbb{N}_{\ell}^{\left(VA3,VA3\right)}\ =\  & 2\,\eta_{\pi}^{2}\,\xi_{\pi}^{4}\,m_{\pi}^{8}\,x(1-x)\,J_{\ell}(d_{1})^{2}\ ,\\
\mathbb{N}_{\ell}^{\left(VA+,VA-\right)}\ =\  & -\eta_{\pi}^{2}\,\xi_{\pi}^{2}\,m_{\pi}^{6}\,p_{\perp}\,\Big(p_{\perp}\,J_{\ell-1}(d_{1})\,J_{\ell+1}(d_{1})-\xi_{\pi}\,m_{\pi}\,J_{\ell}(d_{1})\,J_{\ell-1}(d_{1})\Big)\ ,\\
\mathbb{N}_{\ell}^{\left(VA+,VA3\right)}\ =\  & -2\,\eta_{\pi}^{2}\,\xi_{\pi}^{3}\,m_{\pi}^{7}\,x\,p_{\perp}\,J_{\ell}(d_{1})\,J_{\ell-1}(d_{1})\ ,\\
\mathbb{N}_{\ell}^{\left(VA-,VA3\right)}\ =\  & 2\,\eta_{\pi}^{2}\,\xi_{\pi}^{3}\,m_{\pi}^{7}\,(1-x)\,\Big(p_{\perp}\,J_{\ell+1}(d_{1})\,J_{\ell}(d_{1})-\xi_{\pi}\,m_{\pi}\,J_{\ell}(d_{1})^{2}\Big)\ ,\label{N_VA+_VA3}
\end{align}
with $d_{1}=\xi_{\pi}\,p_{\perp}/(\eta_{\pi}\,m_{\pi}\,x)$. Throughout
the calculation we have used the Bessel function recurrence relation
\begin{equation}
J_{\ell-1}(d_{1})+J_{\ell+1}(d_{1})\ =\ \frac{2\,\ell}{d_{1}}\,J_{\ell}(d_{1})\ .
\end{equation}

Thus, from the above expressions the $\pi^{+}\to l^{+}\nu_{l}$ decay width
is found to be given by
\begin{equation}
\Gamma(\pi^{+}\to l^{+}\nu_{l})\ =\ \frac{1}{2\,E_{q}}\,8\,G_{F}^{2}\,\left|V_{\text{ud}}\right|^{2}\,
\sum_{\Sigma,\Sigma^{\prime}}\, \epsilon_{\Sigma,\Sigma'}
\,\text{Re}\left(f_{\Sigma}\,f_{\Sigma^{\prime}}^{\ast}\right)
\,\mathcal{G}^{(\Sigma,\Sigma^{\prime})}\ ,
\label{gammag}
\end{equation}
where
\begin{align}
\mathcal{G}^{(\Sigma,\Sigma^{\prime})}\  & =\ \frac{1}{4}\sum_{\ell}
\int\frac{dx\,d\theta\,dp_{\perp}^{2}}{(2\pi)^{2}}\;
\delta(p_{\perp}^{2}-\bar{p}_{\perp\ell}^{2})\;\mathbb{N}_{\ell}^{(\Sigma,\Sigma^{\prime})}(x,p_{\perp}) \nonumber \\
 & =\ \frac{1}{8\pi}\sum_{\ell} \int dx\;
 \mathbb{N}_{\ell}^{(\Sigma,\Sigma^{\prime})}(x,\bar{p}_{\perp\ell})\ ,
\label{3.150}
\end{align}
with
\begin{equation}
\bar{p}_{\perp\ell}^{2} \ = \
m_{\pi}^{2}\,(1-x)\,\big[x\,(1+2\,\ell\eta_{\pi})-(1-x)\,\xi_{\pi}^{2}-m_{l}^{2}/m_{\pi}^{2}\big]\ .
\label{plbar}
\end{equation}
In Eq.~(\ref{3.150}) we have used the fact that
$\mathbb{N}_{\ell}^{(\Sigma,\Sigma^{\prime})}$ does not depend on $\theta$.
The integration domain for $x$ is restricted by the conditions
\begin{equation}
0 \ \le \ x \ \le\ 1 ,\qquad\qquad 0 \ \le\ \bar{p}_{\perp\ell}^{\,2}\ ,
\end{equation}
from which one gets
\begin{equation}
\frac{\xi_\pi^2+m_l^2/m_\pi^2}{1+\xi_\pi^2+2\,\ell\eta_\pi} \ \leq \ x \ \leq \ 1\
\end{equation}
and
\begin{equation}
-\,\frac{m_{\pi}^{2}-m_{l}^{2}}{2\,\eta_{\pi}\,m_{\pi}^{2}}\ \leq \ \ell\ < \ \infty\ .
\end{equation}

Notice that, according to the discussion in Sec.~\ref{hadamp}, the
form factors $f_{\Sigma}$ turn out to be in general functions of
$m_{\pi}^{2}$, $\xi_\pi$ and $\eta_{\pi}$.

\newcounter{eraIV}
\renewcommand{\thesection}{\arabic{eraIV}}
\renewcommand{\theequation}{\arabic{eraIV}.\arabic{equation}}
\setcounter{eraIV}{5} \setcounter{equation}{0}

\section{Approximated analytical expressions for relevant laser
parameters}

\label{sec5}

As stated in the previous sections, for the physical situations of interest
the laser frequency is expected to be of the eV order in the lab frame,
which implies that the dimensionless Lorentz-invariant parameter
$\eta_{\pi}$ should be of the order of $10^{-8}$ to $10^{-7}$ for pion
energies $E_{q}\lesssim1$~GeV. Concerning the values of $\xi_{\pi}$, as
discussed in the Introduction we consider a range from low values
$\xi_\pi\sim 10^{-5}$ up to values $\xi_{\pi}\sim{\cal O}(1)$. In this way,
for the cases of interest, the argument
$d_{1}=\xi_{\pi}\,\bar{p}_{\perp\ell}/(\eta_{\pi}\,m_{\pi}\,x)$ of the
Bessel functions in the above defined quantities
$\mathbb{N}_{\ell}^{(\Sigma,\Sigma^{\prime})}$ will be in general very
large, allowing us to approximate Bessel functions through oscillatory
asymptotic expressions. In what follows we work out in this limit the
results for the contributions to the $\pi^+\to l^+\nu_l$ decay width
obtained in the previous section, obtaining approximate analytic expressions
and discussing the corresponding orders of magnitude.

\subsection{Leading order contributions to the decay width}
\label{loc}

Let us analyze the leading order contributions to the $\pi^+\to l^+\nu_l$
decay width for the considered ranges of the parameters $\xi_\pi$ and
$\eta_\pi$. As it will be discussed in Sec.~\ref{sec5.3}, the sums over the
number of exchanged photons $\ell$ in Eq.~(\ref{gammag}) turn out to be
strongly dominated by the contributions of terms for which $\ell\sim {\cal
O}(\xi_\pi\eta_{\pi}^{-1})$ or even larger. This means that in
Eqs.~(\ref{N_A1_A1}) to (\ref{N_VA+_VA3}) we have to deal with Bessel
functions of both large argument and large index. It is convenient to introduce
a new variable
\begin{equation}
z=\frac{d_{1}}{\ell} \ = \ \frac{\xi_{\pi}}{\ell\,\eta_{\pi}}\,\frac{\bar{p}_{\perp\ell}}{x\,m_{\pi}}\ ,
\label{zyz}
\end{equation}
so as to write $J_{\ell}(d_{1})=J_{\ell}(\ell z)$, with $z\sim{\cal
O}(\xi_{\pi}/\left(\ell\,\eta_{\pi}\right))$. Then, the function
$J_{\nu}(\nu\,z)$, for large positive (real) values of $\nu$ and positive
$z$, can be approximated using the Debye asymptotic expansions (see
Eqs.~8.452.1 and 8.453.1 of Ref.~\cite{Gradshteyn:1943cpj})
\begin{align}
J_{\nu}(\nu\,z) & \simeq\sqrt{\frac{1}{2\,\nu\,\pi\,\tanh\alpha}}\,
\exp\left(\nu\,\tanh\alpha-\nu\,\alpha\right)\,
\Big[1+\mathcal{O}\left(\nu^{-1}\right)\Big] &
{\rm if\ }z<1\,,{\rm \ with\ }\cosh\alpha=1/z\ ,
\label{3.90}\\[3mm]
J_{\nu}(\nu\,z) & \simeq\sqrt{\frac{2}{\nu\,\pi\,\tan\beta}}\,
\cos\left(\nu\,\tan\beta-\nu\,\beta-\frac{\pi}{4}\right)\,
\Big[1+\mathcal{O}\left(\nu^{-1}\right)\Big] &
{\rm if\ }z>1\,,{\rm \ with\ }\cos\beta=1/z\ .
\label{3.89}
\end{align}
These relations show that $J_{\ell}(\ell\,z)$ exhibits a sharp exponential
decrease if $z<1$, whereas it is highly oscillating if $z>1$. Now, notice
that the functions $\mathcal{G}^{(\Sigma,\Sigma^{\prime})}$ include an
integral over the variable $x$ (see Eq.~(\ref{3.150})), while the integrands
$\mathbb{N}_{\ell}^{(\Sigma,\Sigma^{\prime})}$ involve products of two
Bessel functions. Given the values of $\ell$ we are interested in, we can
safely neglect the contribution to the integrals that comes from the $z<1$
region and focus on the average value of the product of oscillatory Bessel
functions, approximated by Eq.~(\ref{3.89}), within the $z>1$ region.

If relevant contributions to the sum over $\ell$ are those corresponding to
large values of $\ell$ ($\ell \sim \xi_\pi\eta_\pi^{-1}$ becomes larger than
$10^2$ even for $\xi_\pi\sim 10^{-5}$), the sum can be approximated, at the
lowest order in powers of $\eta_\pi$, by an integral over
$\mathfrak{z}=\ell\eta_{\pi}$, which we can take as a continuous variable.
Thus, we have
\begin{equation}
\mathcal{G}^{(\Sigma,\Sigma^{\prime})}\ \simeq\ \frac{1}{8\pi\eta_{\pi}}
\int d\mathfrak{z}\int dx\:\mathbb{N}_{\ell}^{(\Sigma,\Sigma^{\prime})}\ ,\label{Gcal}
\end{equation}
where the functions $\mathbb{N}_{\ell}^{(\Sigma,\Sigma^{\prime})}$ can be
approximated using the asymptotic expression given in Eq.~(\ref{3.89}). For
now, let us consider the case of positive $\ell$; the case of negative
$\ell$ can then be addressed using the relation
$J_{-\ell}(d_{1})=(-1)^{\ell}\,J_{\ell}(d_{1})$. In the region where $z>1$
we have, in terms of the variables $\mathfrak{z}$ and $z$ (notice that $z$
depends on both $\mathfrak{z}$ and $x$),
\begin{eqnarray}
J_{\ell}(\ell\,z) & \simeq & \sqrt{\frac{2\,\eta_{\pi}}{\mathfrak{z}\,\pi\,\sqrt{z^{2}-1}}}\,
\left\{ \cos\left[\frac{\mathfrak{z}}{\eta_{\pi}}\,\left(\sqrt{z^{2}-1}-\arccos z^{-1}\right)-\frac{\pi}{4}\right]\right\} \ ,
\label{88}\\
J_{\ell\pm1}(\ell\,z) & = & J_{\ell\pm1}\left((\ell\pm1)\,\frac{\ell\,z}{(\ell\pm1)}\right)\nonumber \\
 & \simeq & \sqrt{\frac{2\,\eta_{\pi}}{\mathfrak{z}\,\pi\,\sqrt{z^{2}-1}}}\,\bigg\{\frac{1}{z}\,\cos\left[\frac{\mathfrak{z}}{\eta_{\pi}}\,\left(\sqrt{z^{2}-1}-\arccos z^{-1}\right)-\frac{\pi}{4}\right]\nonumber \\
 &  & \pm\;\frac{\sqrt{z^{2}-1}}{z}\,\sin\left[\frac{\mathfrak{z}}{\eta_{\pi}}\,
\left(\sqrt{z^{2}-1}-\arccos z^{-1}\right)-\frac{\pi}{4}\right]\,\bigg\}\ .
\label{89}
\end{eqnarray}

The integrals over $\mathfrak{z}$ and $x$ in Eq.~(\ref{Gcal})
can now be approximated by averaging the products of the oscillating
Bessel functions, i.e., by replacing
\begin{align}
J_{\ell}(d_{1})^{2}\ \to & \ \frac{\eta_{\pi}}{\pi\,\mathfrak{z}\,\sqrt{z^{2}-1}}\ ,\nonumber \\
J_{\ell}(d_{1})\,J_{\ell\pm1}(d_{1})\ \to & \ \frac{\eta_{\pi}}{\pi\,\mathfrak{z}\,z\,\sqrt{z^{2}-1}}\ ,\nonumber \\
J_{\ell\pm1}(d_{1})^{2}\ \to & \ \frac{\eta_{\pi}}{\pi\,\mathfrak{z}\,\sqrt{z^{2}-1}}\ ,\nonumber \\
J_{\ell-1}(d_{1})\,J_{\ell+1}(d_{1})\ \to & \
\frac{\eta_{\pi}\left(2-z^{2}\right)}{\pi\,\mathfrak{z}\,z^{2}\,\sqrt{z^{2}-1}}\ .
\label{replace-1}
\end{align}
At this point it is advantageous to first perform the integral over
$\mathfrak{z}$. Taking into account the relations (\ref{zyz}) and
(\ref{plbar}), and performing the replacements given in
Eqs.~(\ref{replace-1}), we obtain
\begin{align}
\mathcal{G}^{\left(A1,A1\right)} & \ =\ \frac{1}{8\pi}\int_{\bar{x}}^{1}\,dx\;\frac{1}{2}\,m_{l}^{2}\,(m_{\pi}^{2}-m_{l}^{2})\,I_{0}\ ,\label{app1}\\
\mathcal{G}^{\left(A1,VA\pm\right)} & \ =\ -\frac{1}{8\pi}\int_{\bar{x}}^{1}\,dx\;\,\eta_{\pi}\,m_{\pi}^{4}\,m_{l}^{2}\,x\,I_{1}\ ,\label{G_A1_VA+-}\\
\mathcal{G}^{\left(A1,VA3\right)} & \ =\ \frac{1}{8\pi}\int_{\bar{x}}^{1}\,dx\;2\,\eta_{\pi}\,\xi_{\pi}^{2}\,m_{l}^{2}\,m_{\pi}^{4}\,\left(1-x\right)\,I_{0}\ ,\label{G_A1_VA3}\\
\mathcal{G}^{\left(VA+,VA+\right)} & \ =\ \frac{1}{8\pi}\int_{\bar{x}}^{1}\,dx\;\frac{\eta_{\pi}^{2}}{2}\,\xi_{\pi}^{2}\,m_{\pi}^{8}\,x\,\left[\left(x-\xi_{\pi}^{2}\,(1-x)-m_{l}^{2}/m_{\pi}^{2}\right)\,I_{0}+2\,x\,I_{1}\right]\ ,\\
\mathcal{G}^{\left(VA-,VA-\right)} & \ =\ \frac{1}{8\pi}\int_{\bar{x}}^{1}\,dx\;\frac{\eta_{\pi}^{2}}{2}\,\xi_{\pi}^{2}\,m_{\pi}^{8}\,\left(1-x\right)\,\left[\left(1-x+\xi_{\pi}^{2}\,(2-x)+m_{l}^{2}/m_{\pi}^{2}\right)\,I_{0}-2\,x\,I_{1}\right]\ ,\\
\mathcal{G}^{\left(VA3,VA3\right)} & \ =\ \frac{1}{8\pi}\int_{\bar{x}}^{1}\,dx\;2\eta_{\pi}^{2}\,\xi_{\pi}^{4}\,m_{\pi}^{8}\,x\,(1-x)\,I_{0}\ ,\\
\mathcal{G}^{\left(VA+,VA-\right)} & \ =
\ -\frac{1}{8\pi}\int_{\bar{x}}^{1}\,dx\;\eta_{\pi}^{2}\,m_{\pi}^{8}
\left[-\xi_{\pi}^{2}\,(1-x)\left(x-\xi_{\pi}^{2}\,(1-x)-m_{l}^{2}/m_{\pi}^{2}\right)
I_{0}-\xi_{\pi}^{2}\,(3-2\,x)\,x\,I_{1}+2\,x^{2}I_{2}\right]\ ,\\
\mathcal{G}^{\left(VA+,VA3\right)} & \ =\;-\frac{1}{8\pi}\int_{\bar{x}}^{1}\,dx\;2\,\eta_{\pi}^{2}\,\xi_{\pi}^{2}\,m_{\pi}^{8}\,x^{2}\,I_{1}\ ,\\
\mathcal{G}^{\left(VA-,VA3\right)} & \ =\ \frac{1}{8\pi}\int_{\bar{x}}^{1}\,dx\;2\eta_{\pi}^{2}\,\xi_{\pi}^{2}\,m_{\pi}^{8}\,(1-x)\,(x\,I_{1}-\xi_{\pi}^{2}\,I_{0})\ ,\label{appfin}
\end{align}
where we have defined
\begin{equation}
I_{n}=\frac{1}{\eta_{\pi}}\int_{\mathfrak{z}_{-}}^{\mathfrak{z}_{+}}d\mathfrak{z}\,
\frac{\eta_{\pi}\,
\mathfrak{z}^{n}}{\pi\,\sqrt{(\mathfrak{z}-\mathfrak{z}_{-})\,(\mathfrak{z}_{+}-\mathfrak{z})}}\ .
\label{In_def}
\end{equation}
The integration limits $\bar{x}$ and $\mathfrak{z}_{\pm}$ are determined by
the condition $z^{2}\ge 1$. One has
\begin{equation}
\bar{x}=\frac{m_{l}^{2}}{m_{\pi}^{2}}
\end{equation}
and
\begin{equation}
\mathfrak{z}_{\pm}\ =\;\frac{\xi_{\pi}^{2}\,(1-x)\,\pm\,\xi_{\pi}\,\sqrt{(1-x)\,(x-m_{l}^{2}/m_{\pi}^{2})}}{x}\ .
\label{zpm}
\end{equation}

The integrals $I_{n}$ can be easily performed, obtaining
\begin{equation}
I_{0}=1\ ,\qquad\qquad
I_{1}=\frac{\xi_{\pi}^{2}\,(1-x)}{x}\ ,\qquad\qquad
I_{2}=\frac{\xi_{\pi}^{2}\,(1-x)\,}{2\,x^{2}}\left[x+2\xi_{\pi}^{2}\,(1-x)-m_{l}^{2}/m_{\pi}^{2}\right]\ .
\label{In_res}
\end{equation}
After integration over $x$ we obtain
\begin{eqnarray}
\mathcal{G}^{\left(A1,A1\right)} & \!=\! & \frac{1}{16\pi}\,m_{l}^{2}\,m_{\pi}^{2}\left(1-\frac{m_{l}^{2}}{m_{\pi}^{2}}\right)^{2}\ ,\label{primeraG}\\
\mathcal{G}^{\left(A1,VA+\right)} & \!=\! & \mathcal{G}^{\left(A1,VA-\right)}\ =\
-\,\frac{1}{2}\,\mathcal{G}^{\left(A1,VA3\right)}\ =\ -\,\frac{1}{16\pi}\,\eta_{\pi}\,\xi_{\pi}^{2}\,m_{l}^{2}\,m_{\pi}^{4}\,\left(1-\frac{m_{l}^{2}}{m_{\pi}^{2}}\right)^{2}\ ,
\label{lineal_etapi}\\[2mm]
\mathcal{G}^{\left(VA+,VA+\right)} & \!=\! & \mathcal{G}^{\left(VA-,VA-\right)}\ =\
\frac{1}{96\pi}\,\eta_{\pi}^{2}\,\xi_{\pi}^{2}\,m_{\pi}^{8}\,\left(1-\frac{m_{l}^{2}}{m_{\pi}^{2}}\right)^{2}\,
\left[2+\xi_{\pi}^{2}+(1+2\,\xi_{\pi}^{2})\frac{m_{l}^{2}}{m_{\pi}^{2}}\,\right]\ ,\\[3mm]
\mathcal{G}^{\left(VA+,VA3\right)} & \!=\! & \mathcal{G}^{\left(VA-,VA3\right)}\ =\ -\mathcal{G}^{\left(VA3,VA3\right)}\ =
\ -2\,\mathcal{G}^{\left(VA+,VA-\right)}\nonumber \\[2mm]
 & \!=\! & -\frac{1}{24\,\pi}\,\eta_{\pi}^{2}\,\xi_{\pi}^{4}\,m_{\pi}^{8}\,\left(1-\frac{m_{l}^{2}}{m_{\pi}^{2}}\right)^{2}\,\left(1+\frac{2\,m_{l}^{2}}{m_{\pi}^{2}}\right)\ .\label{ultimaG}
\end{eqnarray}

Finally, replacing in Eq.~(\ref{gammag}), and turning back to the
form factors $f_{V}$, $f_{Ai}$ according to the relations in Eq.~(\ref{deff}),
we arrive at the result
\begin{equation}
\Gamma(\pi^{+}\to l^{+}\nu_{l})\ =\
\frac{G_{F}^{2}\,|V_{\text{ud}}|^{2}}{4\pi\,E_{q}}\;m_{\pi}^{2}\,
\left(1-\frac{m_{l}^{2}}{m_{\pi}^{2}}\right)^{2}\,\sum_{\sigma,\sigma'}
\,\epsilon_{\sigma,\sigma'}\,C_{\sigma,\sigma'}\,
{\rm Re}\big(f_{\sigma}^\ast\,f_{\sigma'}\big)\ ,
\label{mainwidth}
\end{equation}
where $\sigma,\sigma'$ run over the indices $V$ and $Ai$, $i=1,2,3$ and
$\epsilon_{\sigma,\sigma'}$ is defined as in Eq.~(\ref{defepsilon}). At
leading orders in powers of $\eta_{\pi}$, the coefficients
$C_{\sigma,\sigma'}$ are found to be given by
\begin{eqnarray}
C_{A1,A1} & = & m_{l}^{2}\ ,\label{ca1a1}\\
C_{A2,A2} & = & C_{V,V}\ =\ \frac{2}{3}\,\eta_{\pi}^{2}\,\xi_{\pi}^{2}\,m_{\pi}^{6}\,\left(1+\frac{m_{l}^{2}}{2m_{\pi}^{2}}\right)\ ,
\label{ca2a2} \\
C_{A3,A3} & = & \frac{2}{3}\,\eta_{\pi}^{4}\,\xi_{\pi}^{4}\,m_{\pi}^{10}\,\left(1+\frac{2\,m_{l}^{2}}{m_{\pi}^{2}}\right)\ ,\\
C_{A1,\sigma} & \sim & m_{l}^{2}\times{\cal O}(\eta_{\pi}^{1+\gamma_{A1,\sigma}})\qquad{\rm for}\ \sigma=A2,V\ ,
\label{a1v} \\[2mm]
C_{A1,A3} & = & 2\,\eta_{\pi}^{2}\,\xi_{\pi}^{2}\,m_{\pi}^{4}\,m_{l}^{2}\ ,\label{ca1a3}\\[2mm]
C_{A2,V} & \sim & {\cal O}(\eta_{\pi}^{2+\gamma_{A2,V}})\ ,\label{ca2v}\\[2mm]
C_{\sigma,A3} & \sim & {\cal O}(\eta_{\pi}^{3+\gamma_{\sigma,A3}})\qquad{\rm for}\ \sigma=A2,V\ ,
\label{ca2a3_cva3}
\end{eqnarray}
where, in general, $0 < \gamma_{\sigma,\sigma'}\leq 1$.

Some comments regarding these results are in order. Firstly, we point out
that the expressions in Eqs.~(\ref{ca1a1}) to (\ref{ca2a3_cva3}),
corresponding to the terms of the sum in Eq.~(\ref{mainwidth}), are of
different orders in powers of $\eta_{\pi}$.  At the leading order, the width
is given just by the contribution of the $|f_{A1}|^{2}$ term, which is
essentially the result for the width in vacuum, see Eq.~(\ref{gammavac}).
Thus, it is essential to study the next-to-leading order contribution.
Notice that from Eq.~(\ref{lineal_etapi}) one would expect the coefficients
of the cross terms $f_{A1}^\ast\,f_{V}$ and $f_{A1}^\ast\,f_{A2}$ to be of
order $\eta_{\pi}$. Nevertheless, it is seen that these ${\cal O}(\eta_\pi)$
contributions get cancelled: one has
\begin{eqnarray}
C_{A1,V} & \propto & \mathcal{G}^{\left(A1,VA+\right)} + \mathcal{G}^{\left(A1,VA-\right)} +
\mathcal{G}^{\left(A1,VA3\right)}\ , \label{gmm3} \\
C_{A1,A2} & \propto & \mathcal{G}^{\left(A1,VA+\right)} -
\mathcal{G}^{\left(A1,VA-\right)}\ , \label{gmm}
\end{eqnarray}
which are zero according to the relations in Eq.~(\ref{lineal_etapi})  (in
fact, it can be seen that these cancellations occur even before performing
the integral over $x$). The natural question at this point is whether we are
able to determine the dominant contributions to these terms. Although at
first sight one could guess that the lowest nonzero contributions should be
${\cal O}(\eta_{\pi}^{2})$, it can be seen that contributions of ${\cal
O}(\eta_{\pi}^{4/3})$ show up. A detailed analysis of this issue is given in
the next subsection. Notice that cancellations also occur for other
coefficients of higher order in $\eta_\pi$, as stated in Eqs.~(\ref{ca2v})
and (\ref{ca2a3_cva3}).

It is also worth mentioning that the contributions involving $f_{A1}$ vanish
in the limit $m_l=0$. Consequently, in the case of decays to electrons, the
coefficients $C_{A2,A2}$ and $C_{V,V}$ would be competitive for external
fields of very large frequencies or amplitudes. From Eqs.~(\ref{ca1a1}) and
(\ref{ca2a2}), the ratio between the contributions of $|f_{A2}|^2$ and
$|f_{A1}|^2$ terms to the $\pi^+\to e^+\nu_e$ decay width is given by
\begin{equation}
\frac{C_{A2,A2}\,|f_{A2}|^2}{C_{A1,A1}\,|f_{A1}|^2}\
\simeq \
\frac{2\,\eta_{\pi}^{2}\,\xi_{\pi}^{2}\,m_{\pi}^{6}}{3m_e^2}\,
\frac{|f_{A2}|^2}{|f_{A1}|^2}\ \sim \ 5\times 10^4 \,\eta_{\pi}^{2}\,\xi_{\pi}^{2}
\, \frac{m_\pi^4\,|f_{A2}|^2}{|f_{A1}|^2}\ .
\end{equation}
On the other hand, from Refs.~\cite{Coppola:2019uyr,Adhikari:2024vhs}
one can expect the form factors to be of order $|f_{A1}|\sim f_\pi$,
$|f_{A2}|\sim 1/f_\pi$. Hence, one would get sizeable corrections for values
of $\eta_\pi\xi_\pi\sim {\cal O}(10^{-2})$. A similar situation occurs for
the $|f_{V}|^2$ contribution.

\subsection{Dominant subleading contribution in powers of
$\eta_\pi$}

As stated, the dominant subleading contribution to the total $\pi^+\to
l^+\nu_l$ width is expected to be given by the cross terms
$f_{A1}^\ast\,f_{V}$ and $f_{A1}^\ast\,f_{A2}$. The estimation of the
corresponding coefficients requires some detailed analysis of the behavior
of Bessel functions $J_\nu(\nu z)$ for large $\nu$ in the region $z\simeq
1$, which represents a complex but well studied problem~\cite{DebyeJen}.

\begin{figure}[htb]
\centering{}
\vspace{0.5cm}
\includegraphics[width=0.9\textwidth]{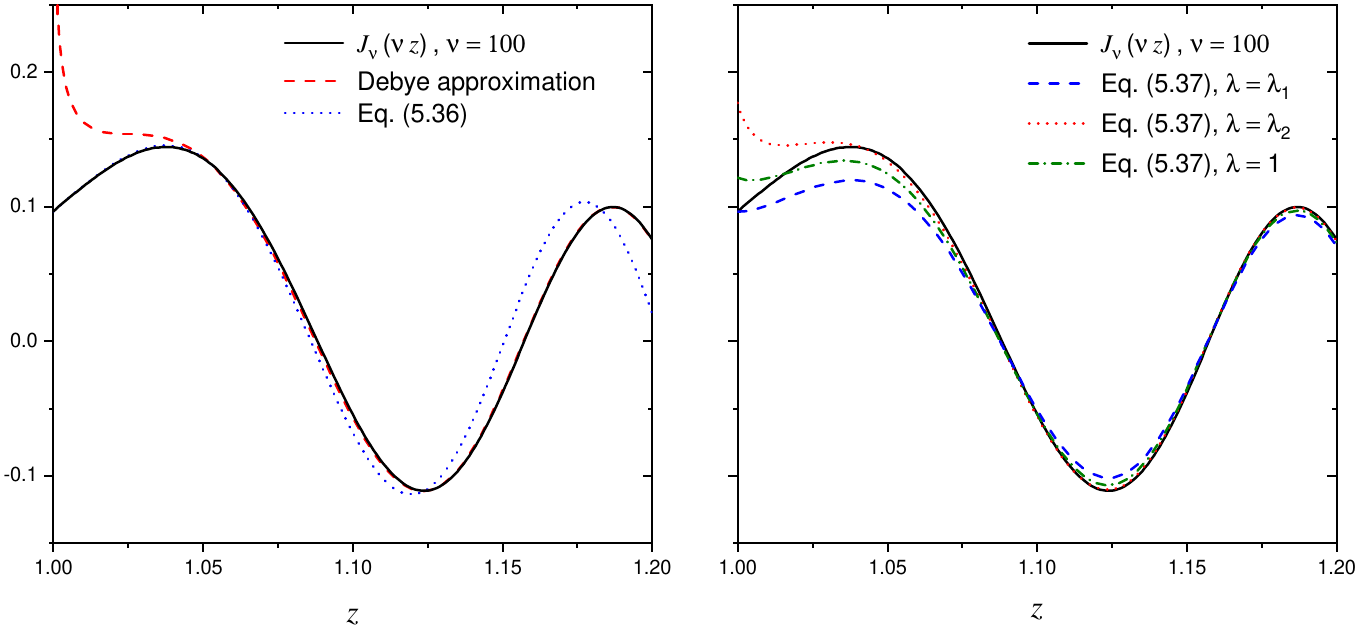}
\caption{Bessel function $J_\nu(\nu z)$ and approximate expressions for
$\nu=100$. The black curves in both panels correspond to the function
$J_{\nu}\left(\nu\,z\right)$. In the left panel, the red dashed line
corresponds to the Debye approximation given by Eq.~(\ref{88}), whereas the
blue dotted line corresponds to the expansion in the $z\sim 1$ region,
Eq.~(\ref{D.10}). In the right panel, the dashed, dotted and dashed
dotted lines correspond to the regularized approximation in
Eq.~(\ref{Bessel_reg0}) for $\lambda = \lambda_1$, $\lambda =\lambda_2$ and
$\lambda =1$, respectively.}
\label{FigBessel100}
\end{figure}

The Debye asymptotic expansions, contrary to its appearance in
Eqs.~(\ref{3.90}) and (\ref{3.89}), are not expansions in inverse powers of
the large index $\nu$, but turn out to be uniform expansions in inverse
powers of the large parameter $\nu^{2/3}\,(z-1)/z$ for $z>1$ and the large
parameter $\nu^{2/3}\,$(1$-z)$ for $z<1$~\cite{DebyeMat}. Fractional powers
of $\nu$ arise, in particular, when one approaches the value $z=1$. In
fact, for values of $z$ around 1, both Debye asymptotic expansions fail.
In the transition region one can instead use the expansion
\begin{equation}
J_{\nu}\left(\nu\,\big(1+a\,\nu^{-2/3}\big)\right) \
= \ \frac{2^{1/3}}{\nu^{1/3}}\,
\text{Ai}\big(-2^{1/3}a\big)\,\left[1+\mathcal{O}\left(\nu^{-2/3}\right)\right]\ ,
\label{D.10}
\end{equation}
where ${\rm Ai}(x)$ is the Airy function. In the left panel of
Fig.~\ref{FigBessel100} we show, for $\nu = 100$, the behaviors of the exact
$J_{\nu}(\nu\,z)$ function (solid black line), the Debye expansion given by
Eq.~(\ref{88}) (red dashed line) and the approximated result for the
$z\simeq 1$ in Eq.~(\ref{D.10}) (blue dotted line). It is seen that Debye
expansion fails in the region of $z$ close to 1, where it diverges. Though
this divergence is smooth enough to ensure that the integral over $z$ is
finite, its effects are worth to be taken into account.

The unwanted divergence at $z=1$ can be regularized by introducing a cutoff
$\lambda$ in the Debye approximation. Here we propose to implement such a
regularization by considering the approximate expression
\begin{equation}
J_{\nu}(\nu\,z)\ \simeq \
\sqrt{\frac{2}{\nu\,\pi\,\sqrt{z^{2}-1+\lambda\,\nu^{-2/3}}}}\;
\cos\left[\nu\,\left(\sqrt{z^{2}-1}-\arccos z^{-1}\right)-\frac{\pi}{4}\right] \ .
\label{Bessel_reg0}
\end{equation}
One should now choose some criterion to determine an appropriate value for
$\lambda$. A simple option is to impose coincidence between the regularized
function and the exact Bessel function at $z=1$. This implies to take
\begin{equation}
\lambda \ = \ \lambda_{1} \ = \
\frac{1}{\Big[2^{1/3}\sqrt{\pi}\,\text{Ai}(0)\Big]^{4}}\ = \ 2.531\ .
\end{equation}
A second possibility is to impose coincidence between the regularized
function and the exact Bessel function just at the first maximum. The
position of this maximum is defined by the condition
$\sqrt{z^{2}-1}-\arccos(1/z)\simeq \pi/(4\nu)$, which implies $z \simeq
1+[3\,\pi/(8\,\sqrt{2})]^{2/3}\,\nu^{-2/3}$. Thus, this criterion leads to
\begin{equation}
\lambda \ = \
\lambda_{2} \ = \
\frac{2^{2/3}}{\Big[\sqrt{\pi}\,
\text{Ai}(-(3\pi/8)^{2/3})\Big]^{4}}\,-\,
\bigg(\frac{3\,\pi}{4}\bigg)^{2/3} =\ 0.222\ .
\end{equation}
The behavior of the approximate expression in Eq.~(\ref{Bessel_reg0})
for different values of $\lambda$ is illustrated in the right panel of
Fig.~\ref{FigBessel100}. It can be seen that the choice
$\lambda=\lambda_{1}$ (blue dashed line) underestimates the Bessel function,
whereas the choice $\lambda=\lambda_{2}$ (red dotted line) overestimates it.
Since our results for the functions $\mathcal{G}^{(A1,VA\pm)}$ and
$\mathcal{G}^{(A1,VA3)}$ ultimately depend on $\sqrt{\lambda}$ (see below),
an intermediate value can be obtained from
$\sqrt{\lambda}=\left(\sqrt{\lambda_{1}}+\sqrt{\lambda_{2}}\right)/2$. The
result can be approximated by $\lambda=1$, which corresponds to the green
dotted-dashed line in the right panel of Fig.~\ref{FigBessel100}. Taking a
given value of $\lambda$, we have checked that the regularization
method is stable regarding the effect of next-to-leading order corrections
to the approximated expressions for Bessel functions.

The introduction of the regularized approximate function in
Eq.~(\ref{Bessel_reg0}) leads to corrections to our results that can be
readily evaluated. Essentially, the integrals $I_{n}$, $n=0,1,2$, defined in
Eq.~(\ref{In_def}) have to be replaced by regularized integrals given by
\begin{equation}
I_{n}^{\text{r}}\ =\ \frac{1}{\eta_{\pi}}\int_{\mathfrak{z}_{-}}^{\mathfrak{z}_{+}}d\mathfrak{z}\,
\frac{\eta_{\pi}\,\mathfrak{z}^{n}}{\pi\,\sqrt{(\mathfrak{z}-\mathfrak{z}_{-})\,
(\mathfrak{z}_{+}-\mathfrak{z})+\lambda\,\eta_{\pi}^{2/3}\mathfrak{z}^{4/3}}}\ .
\label{Ireg}
\end{equation}
As shown in App.~\ref{AppB}, at the leading order in $\eta_\pi$ one obtains
\begin{equation}
I_{n}^{\text{r}} - I_{n} \ = \
-\frac{2}{\pi}\,\eta_{\pi}^{1/3}\,\sqrt{\lambda}\;
\frac{1}{\left(\mathfrak{z}_{+}-\mathfrak{z}_{-}\right)}\,\left(\mathfrak{z}_{-}^{n+2/3}+\mathfrak{z}_{+}^{n+2/3}\right)\
, \label{Inreg_res}
\end{equation}
where $I_{n}$, $n=0,1,2$, stand for our previous results, given in Eq.~(\ref{In_res}).

Now we can go back to Eqs.~(\ref{G_A1_VA+-}) and (\ref{G_A1_VA3}). After
integration over $x$, the above expressions lead to
\begin{align}
\mathcal{G}^{\left(A1,VA\pm\right)} & \ =\ -\frac{1}{8\pi}\,\eta_{\pi}\,m_{\pi}^{4}\,m_{l}^{2}\,
\left[\frac{1}{2}\,\xi_{\pi}^{2}\,\left(1-\bar{x}\right)^{2}-
\left(1-\bar x\right)^{2}\,\left(1+2\,\bar x\right)\,\eta_{\pi}^{1/3}\,\frac{1}{\pi}\,\sqrt{\lambda}\,\xi_{\pi}^{5/3}\,f(\xi_{\pi},\bar{x})\right]\ ,\\
\mathcal{G}^{\left(A1,VA3\right)} & \ =\ \frac{1}{8\pi}2\,\eta_{\pi}\,m_{l}^{2}\,m_{\pi}^{4}\,\left[
\frac{1}{2}\,\xi_{\pi}^{2}\,\left(1-\bar{x}\right)^{2}-\left(1-\bar x\right)^{2}\,
\left(1+2\,\bar x\right)\,\eta_{\pi}^{1/3}\,\frac{1}{\pi}\,\sqrt{\lambda}\,\xi_{\pi}^{5/3}\,g(\xi_{\pi},\bar{x})\right]\ ,
\end{align}
where
\begin{eqnarray}
f(\xi_{\pi},\bar{x}) & = & \frac{1}{\xi_{\pi}^{8/3}\,\left(1-\bar{x}\right)^{2}\,\left(1+2\,\bar{x}\right)}
\int_{\bar{x}}^{1}dx\ \frac{x^{2}}{\sqrt{(1-x)(x-\bar{x})}}\left(\mathfrak{z}_{+}^{5/3}+\mathfrak{z}_{-}^{5/3}\right)\ , \\
g(\xi_{\pi},\bar{x}) & = & \frac{1}{\xi_{\pi}^{2/3}\,\left(1-\bar{x}\right)^{2}\,\left(1+2\,\bar{x}\right)}
\int_{\bar{x}}^{1}dx\
\frac{x\sqrt{1-x}}{\sqrt{x-\bar{x}}}\left(\mathfrak{z}_{+}^{2/3}+\mathfrak{z}_{-}^{2/3}\right)\ .
\end{eqnarray}
It can be seen that the combination $\mathcal{G}^{\left(A1,VA+\right)} +
\mathcal{G}^{\left(A1,VA-\right)} + \mathcal{G}^{\left(A1,VA3\right)}$ in
Eq.~(\ref{gmm3}) is now different from zero. Hence, we get a subleading
contribution to the coefficient $C_{A1,V}$ given by
\begin{equation}
C_{A1,V}=\frac{4}{\pi}\,\eta_{\pi}^{4/3}\,m_{\pi}^{2}\,m_{l}^{2}\,
\left(1+2\,\frac{m_{l}^{2}}{m_{\pi}^{2}}\right)\,\xi_{\pi}^{5/3}\,\sqrt{\lambda}\,
\big[f(\xi_{\pi},\bar{x})-g(\xi_{\pi},\bar{x})\big]\ ,
\label{ca1vapp}
\end{equation}
i.e., the power of $\eta_\pi$ for the cross term $f_{A1}^\ast\,f_{V}$ in
Eq.~(\ref{a1v}) is given by $1+\gamma_{A1,V}$ with $\gamma_{A1,V} = 1/3$. On
the other hand, the cancellation between $\mathcal{G}^{\left(A1,VA+\right)}
- \mathcal{G}^{\left(A1,VA-\right)}$ still holds valid, leading to
$\gamma_{A1,A2} \geq 1$ for the coefficient of the
$f_{A1}^\ast\,f_{A2}$ term. We conclude from this analysis that in the
limit $\eta_\pi\ll 1$ the contribution from the cross term
$f_{A1}^\ast\,f_{V}$, of order $\eta_{\pi}^{4/3}$, is the dominant one
beyond the leading $|f_{A1}|^2$ term.

The above conclusions do not vary qualitatively with the chosen value of
$\lambda$, which, as stated, should be taken to be of order unity in order
to get a good approximation to the exact Bessel functions. Moreover, the
functions $f(\xi_{\pi},\bar{x})$ and $g(\xi_{\pi},\bar{x})$ have been
defined in such a way that they vary just smoothly for values of $\xi_\pi$
within a range from $10^{-5}$ to 1. This is illustrated in
Table~\ref{Table1}, where numerical values of both functions, as well as the
difference $f(\xi_{\pi},\bar{x})- g(\xi_{\pi},\bar{x})$, are quoted for
$\xi_\pi = 10^{-5}$ and $\xi_\pi=1$. In particular, in the limit $\xi_\pi\ll
1$, using the change of variables $x=y\left(1-\bar{x}\right)+\bar{x}$ the
integrals can be simplified to
\begin{eqnarray}
f(\xi_{\pi},\bar{x}) & \simeq & \frac{10}{3\,\left(1+2\,\bar{x}\right)}
\int_{0}^{1}dy\,\left(y+\frac{\bar{x}}{\left(1-\bar{x}\right)}\right)^{1/3}\,
\frac{\left(1-y\right)^{5/6}}{y^{1/6}} \ ,\nonumber
\\[3mm]
g(\xi_{\pi},\bar{x}) & \simeq & \frac{2}{\left(1+2\,\bar{x}\right)}\int_{0}^{1}dy\,
\left(y+\frac{\bar{x}}{\left(1-\bar{x}\right)}\right)^{1/3}\,
 \frac{\left(1-y\right)^{5/6}}{y^{1/6}} \ .
\end{eqnarray}
\begin{table}[hbt]
\begin{centering}
\begin{tabular}{c|c|c|c}
 & $f(\xi_{\pi},\bar{x})$ & $g(\xi_{\pi},\bar{x})$ & $f(\xi_{\pi},\bar{x})-g(\xi_{\pi},\bar{x})$
\tabularnewline[1mm]
\hline
$\pi^{+}\rightarrow e^{+}\,\bar{\nu}_{e}\quad$
\begin{tabular}{l}
$\xi_{\pi}=10^{-5}$\tabularnewline
$\xi_{\pi}=1$\tabularnewline
\end{tabular} & %
\begin{tabular}{c}
$1.454$\tabularnewline
$1.630$\tabularnewline
\end{tabular} & %
\begin{tabular}{c}
$0.873$\tabularnewline
$1.227$\tabularnewline
\end{tabular} & %
\begin{tabular}{c}
$0.582$\tabularnewline
$0.404$\tabularnewline
\end{tabular}\tabularnewline
\hline
$\pi^{+}\rightarrow\mu^{+}\,\bar{\nu}_{\mu}\quad$%
\begin{tabular}{l}
$\xi_{\pi}=10^{-5}$\tabularnewline
$\xi_{\pi}=1$\tabularnewline
\end{tabular} & %
\begin{tabular}{c}
$1.294$\tabularnewline
$1.669$\tabularnewline
\end{tabular} & %
\begin{tabular}{c}
$0.776$\tabularnewline
$1.362$\tabularnewline
\end{tabular} & %
\begin{tabular}{c}
$0.517$\tabularnewline
$0.307$\tabularnewline
\end{tabular}\tabularnewline
\hline
\end{tabular}
\par\end{centering}
\caption{Values of the functions $f(\xi_{\pi},\bar{x})$,
$g(\xi_{\pi},\bar{x})$ and $f(\xi_{\pi},\bar{x})-g(\xi_{\pi},\bar{x})$ for
$\pi^+\to e^+\nu_e$ and $\pi^+\to \mu^+\nu_\mu$ decays, considering
different values of $\xi_{\pi}$.}
\label{Table1}
\end{table}

According to the above estimations for the functions $f(\xi_{\pi},\bar{x})$
and $g(\xi_{\pi},\bar{x})$, the corrections to the dominant contribution are
found to be of the same order of magnitude for the decays into positrons and
muons. We obtain
\begin{equation}
\frac{C_{A1,V}}{C_{A1,A1}}\ \sim \
\frac{4}{\pi}\,m_{\pi}^{2}\,\eta_{\pi}^{4/3}\,\xi_{\pi}^{5/3}\,r_l(\xi_\pi,\bar x)\,\sqrt{\lambda}\ ,
\end{equation}
where $\sqrt{\lambda}$ is expected to lie within a range from $\sim 0.5$ to
1.6, while $r_l(\xi_\pi,\bar x)\simeq 0.5$ for $l=e^+$ and $r_l(\xi_\pi,\bar
x) \simeq 1$ for $l=\mu^+$.

\subsection{Contributions from relative low values of $\ell$}

\label{sec5.3}

In our calculation we have assumed that the main contributions to the decay
width correspond to values of $|\ell|$ of ${\cal O}(\xi_\pi\eta_{\pi}^{-1})$
or even larger. Let us now consider the contributions corresponding to a
relatively low number of exchanged photons, i.e., to $|\ell|\le\ell_0$, with
$\ell_0\,\eta_{\pi}\ll \xi_\pi$. We recall that we are interested in
physical situations for which $\eta_\pi$ is of order $10^{-7}$ to $10^{-8}$,
while the considered values of $\xi_\pi$ range from $10^{-5}$ to 1. Since,
as discussed in Sec.~\ref{loc}, the argument $d_{1}$ of the Bessel functions
in Eqs.~(\ref{N_A1_A1}) to (\ref{N_VA+_VA3}) will be in general a large
number, the arguments of the integrals over $x$ in Eq.~(\ref{3.150}) will
still be highly oscillatory. Therefore, we can approximate the Bessel
functions taking into account the asymptotic behavior~\cite{NIST1}
\begin{equation}
J_{\ell}(d_{1})\ \simeq \ \sqrt{\frac{2}{\pi\,d_{1}}}\,\cos\Big(d_{1}-\ell\,\frac{\pi}{2}-\frac{\pi}{4}\Big)\ ,
\qquad\ell \leq \ell_{0}\;.
\label{BessAsymp1-1}
\end{equation}
In fact, for these small $\ell$ contributions we can still consider the
Debye approximation used in our calculation, Eq.~(\ref{88}), in the limit in
which $z^{2}$ is much larger than one. Hence, the integrals in
Eq.~(\ref{3.150}) can be estimated by introducing in Eqs.~(\ref{N_A1_A1}) to
(\ref{N_VA+_VA3}) similar replacements as in Eqs.~(\ref{replace-1}), taking
$z^{2}=\left(d_{1}/\ell\right)^{2}\gg1$:
\begin{align}
J_{\ell}(d_{1})^{2}\ \to & \ \frac{1}{\pi\,d_{1}}\ ,\nonumber \\
J_{\ell}(d_{1})\,J_{\ell\pm1}(d_{1})\ \to & \ \frac{\ell}{\pi\,d_{1}^{2}}\ \sim\ 0 \ ,\nonumber \\
J_{\ell\pm1}(d_{1})^{2}\ \to & \ \frac{1}{\pi\,d_{1}}\ ,\nonumber \\
J_{\ell-1}(d_{1})\,J_{\ell+1}(d_{1})\ \to & \ -\frac{1}{\pi\,d_{1}}\ .
\label{replace-1-1}
\end{align}

For definiteness, let us consider the case of
$\mathcal{G}^{\left(A1,A1\right)}$, denoting by
$\mathcal{G}^{\left(A1,A1\right)}_{\ell_0}$ the total contribution to
$\mathcal{G}^{\left(A1,A1\right)}$ arising from terms with $|\ell|\leq
\ell_0$. After the above replacements, at the leading order in powers of
$\eta_{\pi}$ we obtain
\begin{equation}
\mathcal{G}^{\left(A1,A1\right)}_{\ell_0}\ = \
\sum_{\ell=-\ell_{0}}^{\ell_{0}}\,\frac{1}{8\pi}\int_{\bar{x}_{0}}^{1}\,dx\;\frac{\eta_{\pi}}{2\pi\,\xi_{\pi}}\,m_{\pi}\,m_{l}^{2}\,(m_{\pi}^{2}-m_{l}^{2})\,\frac{x}{\bar{p}_{\perp0}}\ ,\label{upsa1a1}
\end{equation}
where
\begin{equation}
\bar{x}_{0}=\frac{\xi_{\pi}^{2}+m_{l}^{2}/m_{\pi}^{2}}{1+\xi_{\pi}^{2}}\ ,\qquad\qquad\bar{p}_{\perp0}^{2}=m_{\pi}^{2}\,(1+\xi_{\pi}^{2})\,(1-x)\,(x-\bar{x}_{0})\ .\label{xbar-1}
\end{equation}
This integral can be analytically performed, leading to
\begin{align}
\mathcal{G}^{\left(A1,A1\right)}_{\ell_0}\  & =\ \sum_{\ell=-\ell_{0}}^{\ell_{0}}\,
\frac{1}{32\,\pi}\,\frac{\eta_{\pi}}{\xi_{\pi}\,\left(1+\xi_{\pi}^{2}\right)^{3/2}}\,
m_{l}^{2}\,\left(1-\frac{m_{l}^{2}}{m_{\pi}^{2}}\right)\,\left[\left(1+2\,\xi_{\pi}^{2}\right)
\,m_{\pi}^{2}+m_{l}^{2}\right]\ .
\label{upsa1a1-1}
\end{align}
Notice that the expression in Eq.~(\ref{upsa1a1-1}) is ${\cal
O}(\eta_{\pi}\xi_\pi^{-1})$, and the terms in the sum do not depend on $\ell$.
Therefore, the full contribution of the sum turns out to be negligible when
compared with the total value of $\mathcal{G}^{\left(A1,A1\right)}$, which,
as shown in the previous subsections, is expected to be ${\cal
O}(\eta_{\pi}^{0})$. In other words, the sum of contributions of the form
given in Eq.~(\ref{upsa1a1-1}) will be negligible as long as $\ell$ does not
reach values of order $\xi_\pi\eta_{\pi}^{-1}$. A similar reasoning is shown
to be valid for all contributions to
$\mathcal{G}^{(\Sigma,\Sigma^{\prime})}$ satisfying
$\ell\,\eta_{\pi}\ll\xi_\pi$.

\subsection{Bounds for the number of exchanged photons}

As stated in Sec.~\ref{loc}, for the computation of the total width we have found it
convenient to perform firstly the sum (approximated by an integral) over the
number of exchanged photons and then the integral over the phase space
variable $x$. However, it is also interesting to consider the alternative
procedure, i.e., to integrate over $x$ for a given value of $\ell$, as
expressed in Eq.~(\ref{3.150}).

Since the usage of the Debye expansion of Bessel functions discussed in
Sec.~\ref{loc} is always justified, the substitutions in
Eqs.~(\ref{replace-1}) can still be introduced, and once again the phase
space region turns out to be constrained by the condition $z^{2}>1$. For a
given $\ell$, the integral over $x=p^{-}/q^{-}$ is shown to be limited to
$x_{-}<x<x_{+}$, where
\begin{equation}
x_{\pm}\ =\ \frac{1+2\,\ell\,\eta_{\pi}+\xi_{\pi}^{2}\,(1+\Delta^{2})\,\pm\,
\sqrt{(1+2\,\ell\,\eta_{\pi}-m_{l}^{2}/m_{\pi}^{2})^{2}-4\,\ell^{2}\,\eta_{\pi}^{2}\,\Delta^{2}}}
{2\,\Big[1+(\ell\,\eta_{\pi}+\xi_{\pi}^{2})^{2}/\xi_{\pi}^{2}\Big]}\ ,
\end{equation}
with
\begin{equation}
\Delta\ =\ \sqrt{\,1\,+\,\frac{m_{l}^{2}}{\xi_{\pi}^{2}\,m_{\pi}^{2}}}\ .
\end{equation}
In turn, the sum over $\ell$ is found to be restricted to
$\ell_{\min}<\ell<\ell_{{\rm max}}$, where
\begin{equation}
\ell_{{\rm min}}=-\,\frac{1-m_{l}^{2}/m_{\pi}^{2}}{2\,\eta_{\pi}(\Delta+1)}\ ,\qquad\qquad\ell_{{\rm max}}=
\frac{1-m_{l}^{2}/m_{\pi}^{2}}{2\,\eta_{\pi}(\Delta-1)}\ .
\label{lminmax}
\end{equation}
That is to say, one finds upper and lower cutoffs for the number of photons to be
exchanged with the electromagnetic field.

The existence of these bounds has been observed from numerical calculations
in Ref.~\cite{Mouslih:2020pfd}. In that reference the authors consider
charged pion decays to muons, taking values of $\eta_\pi$ similar to those
considered in our work and values of $\xi_{\pi}$ of order $10^{-6}$ to
$10^{-5}$, as well as some kinematical cuts in the muon momentum. It can be
seen, indeed, that the obtained numerical results are consistent with the
approximate limits $\ell \simeq
\pm\xi_{\pi}(1-m_{l}^{2}/m_{\pi}^{2})\,m_{\pi}/(2\eta_{\pi}m_{l})$ that
follow from Eqs.~(\ref{lminmax}) for such low values of $\xi_{\pi}$. On the
basis of these numerical calculations, after summation over $\ell$, the
analysis in Ref.~\cite{Mouslih:2020pfd} concludes that the
$\pi^+\to\mu^+\nu_\mu$ decay width is largely dominated by the contribution
of the $|f_{A1}|^{2}$ term, which, as stated, is essentially the result for
the width in vacuum. Notice, however, that the feasibility of the numerical
analysis is favored by the low value of $\xi_\pi$, which restricts the
number of exchanged photons to values that does not exceed the order of
$10^3$ (for example, taking $\eta_\pi\sim 10^{-8}$ and $\xi_\pi\sim
10^{-5}$, from Eqs.~(\ref{lminmax}) one gets $\ell_{{\rm max}}\simeq
|\ell_{{\rm min}}|\sim 300$). In our work, the same conclusion can be
obtained from the analytical expressions in Eqs.~(\ref{mainwidth}) to
(\ref{ca2a3_cva3}); this allows us to extend its validity to values of
$\xi_\pi$ up of order unity, which could be reached by present experimental
facilities. For $\eta_\pi\simeq 10^{-8}$ and $\xi_\pi\simeq 1$,
according to Eqs.~(\ref{lminmax}) the maximum number of exchanged photons
becomes enhanced up to $\simeq 3\times 10^7$, challenging computational
capabilities. Moreover, from our analytical calculation it is seen that the
fact that the width is dominated by the $|f_{A1}|^{2}$ term also holds in
the case of the decays to electrons, for which upper and lower limits for
the number of exchanged photons reach values of $10^4$ already for
$\xi_\pi\sim 10^{-6}$. Concerning the bounds for the number of exchanged
photons, it is also worth noticing that for values of $\xi_{\pi}$ of order
unity upper and lower limits of $\ell$ are not symmetric. In particular, in
the case of electrons $\Delta-1$ gets suppressed, and consequently the upper
limit for $\ell$ may become enhanced up to values of ${\cal O}(10^{13})$.

\subsection{Connection with the Kroll-Watson formula}

At this point it is interesting to observe that there is a connection
between our expression for the $\pi^+\to l^+\nu_l$ decay width and the
Kroll-Watson formula~\cite{Kroll-Watson}. The latter corresponds to the
differential cross section, in the nonrelativistic limit, for the
scattering of a charged particle by some scattering potential in the presence
of a strong external electromagnetic wave, considering the exchange
of a given number of photons.

Let $\vec p_{0}$ and $\vec p(\ell)$ be the initial and final momenta
of the charged particle, respectively, $\ell$ being the number of emitted
($\ell<0$) or absorbed ($\ell>0$) photons. The Kroll-Watson formula states
that the differential cross section $d\sigma_\ell(p(\ell),p_{0})/d\Omega$
can be written as
\begin{equation}
\frac{d\sigma_{\ell}\big(p(\ell),p_{0}\big)}{d\Omega} \ = \
\frac{|\vec p\left(\ell\right)|}{|\vec p_{0}|}\,
J_{\ell}(x_{\text{KW}})^2\, \frac{d\sigma_{\text{el}}}{d\Omega}\ ,
\label{K-W}
\end{equation}
where $d\sigma_{\text{el}}/d\Omega$ stands for the differential
elastic scattering cross section in the absence of the electromagnetic
field, and $x_{\text{KW}}$ is defined as
\begin{equation}
x_{\text{KW}}\ = \ -Q\,\frac{\vec{A}\cdot
\left(\vec{p}\left(\ell\right)-\vec{p}_{0}\right)}{m\,\omega}\ ,
\end{equation}
where $Q$ is the particle charge and $\vec A$ and $\omega$ are the
parameters that characterize the electromagnetic wave. According to
Ref.~\cite{Kroll-Watson}, Eq.~(\ref{K-W}) is valid when the scattering
potential is weak or when the wave frequency is small, which is given by the
condition
\begin{equation}
\left|\frac{\ell\,\omega\,m}{Q\,\vec{A}\cdot\left(\vec{p}\left(\ell\right)-\vec{p}_{0}\right)}\right|
\ = \ \left|\frac{\ell}{x_{\text{KW}}}\right| \ <\ 1\ .
\end{equation}

Although Ref.~\cite{Kroll-Watson} deals with a scattering problem, the
result in Eq.~(\ref{K-W}) can be connected with our result for the charged
pion decay. First of all, it can be seen that  with our notation and
conventions the quantity $x_{\text{KW}}$ corresponds to our $d_{1}$,
defined as $d_1=e\,a\,|\ABK\cdot p|/(k\cdot p)$ in Eq.~(\ref{d1_d2_0}).
We can take Eq.~(\ref{N_A1_A1}) and perform the replacements
(see Eqs.~(\ref{replace-1}))
\begin{align}
J_{\ell}(d_{1})\,J_{\ell\pm1}(d_{1}) & \ \to\
\frac{\ell}{d_1}\,J_{\ell}(d_{1})^{2}\ ,
\nonumber \\
J_{\ell+1}(d_{1})^2 & \ \to\ J_{\ell}(d_{1})^{2}\ ,
\end{align}
to obtain
\begin{equation}
\mathbb{N}_{\ell}^{\left(A1,A1\right)}\ =\
\frac{m_{l}^{2}}{2} \left(m_{\pi}^{2}-m_{l}^{2}\right)
J_{\ell}\left(d_{1}\right)^{2}\ .
\end{equation}
Thus, identifying the form factor $f_{A1}$ with the pion decay
constant $f_\pi$, at the leading order ${\cal O}(\eta_\pi^0)$ the
differential width can be written as $d\Gamma/dx =
\sum_{\ell}\,d\Gamma_{\ell}/dx$, with
\begin{align}
\frac{d\,\Gamma_{\ell}}{dx} & \ = \
\frac{1}{4\pi\,E_{q}}\;G_{F}^{2}\,V_{\text{ud}}^{2}\,m_{l}^{2}\,(m_{\pi}^{2}-m_{l}^{2})\,
f_{\pi}^{2}\,J_{\ell}(d_{1})^{2} \nonumber \\
 & \ = \ \frac{d\,\Gamma^{(0)}}{dx}\,J_{\ell}(d_{1})^{2}\ ,
 \label{K-W-1}
\end{align}
where $d\,\Gamma^{(0)}/dx=G_{F}^{2} V_{\text{ud}}^{2}
m_{l}^{2}(m_{\pi}^{2}-m_{l}^{2})f_{\pi}^{2}/(4\pi E_{q})$ is the
differential width in absence of the external field. In this way, our result
turns out to be analogous to the Kroll-Watson formula. The factor $|\vec
p(\ell)|/|\vec p_{0}|$ in Eq.~(\ref{K-W}) is related to the fact that  for
the scattering process the differential cross section is defined with
respect to the solid angle, while in our case the differential decay width
is defined with respect to the variable $x=p^{-}/q^{-}$.

Finally, notice that the result given in Eq.~(\ref{K-W-1}) is valid for
``small $\ell$'' (i.e., $|\ell|<|x_{\text{KW}}|$), but also for large values
of $\ell,$ due to the fact that  one is dealing with a weak potential. In
this way, at order $\eta_\pi^0$ we can evaluate the differential
width $d\Gamma/dx$ by performing the sum
\begin{align}
\frac{d\,\Gamma}{dx}  & \ = \ \sum_{\ell=-\infty}^\infty\,\frac{d\,\Gamma_{\ell}}{dx}
\ = \ \frac{d\,\Gamma^{(0)}}{dx}\,\sum_{\ell} \, J_{\ell}(d_{1})^{2}\ = \
\frac{d\,\Gamma^{(0)}}{dx} \ .
\end{align}
Here we have used the relation $\sum_\ell J_\ell(d_1)^2 = I_0 = 1$,
obtained in Sec.~\ref{loc} using the asymptotic expressions for Bessel
functions. After integration over $x$, one recovers the result for
the total width in absence of the external field, Eq.~(\ref{gammavac}).

\newcounter{eraV}
\renewcommand{\thesection}{\arabic{eraV}}
\renewcommand{\theequation}{\arabic{eraV}.\arabic{equation}}
\setcounter{eraV}{6} \setcounter{equation}{0}

\section{Summary and conclusions}

We have obtained analytical expressions for the $\pi^+\to l^+\nu_l$ decay
width in the presence of a background electromagnetic plane wave with
circular polarization. Our first step has been to determine the number of
form factors involved in the decay process. It is shown that for an
electromagnetic plane wave background the hadronic $V-A$ pion-to-vacuum
amplitude can be written in terms of four form factors, three of them
corresponding to the axial-vector current and one to the vector current.
This result coincides with the one obtained in Ref.~\cite{Coppola:2018ygv}
for the case of a background given by a static uniform magnetic field. We
have also shown that, in general, the form factors are functions of the pion
mass and the Lorentz-invariant quantities $\xi_{\pi}$ and $\eta_{\pi}$,
which are related to the amplitude and the frequency of the external wave,
respectively.

We have seen that the most adequate kinematic variables for the description
of the decay process are the fraction of the light front momentum of the
pion ported by the charged lepton, $x=p^{-}/q^{-},$ and the number of involved
photons, $\ell$. In terms of these quantities we have obtained the full
analytic expression for the decay width, given by Eq.~(\ref{gammag}). As
discussed in Sec.~\ref{sec5}, this expression can be approximated taking
into account the values of the parameters related to the external field for
the physical situations of interest, namely, $\eta_\pi\sim 10^{-8}$ to
$10^{-7}$ and $\xi_\pi\sim 10^{-5}$ to 1. Transforming the sum over $\ell$
into an integral and expanding in powers of $\eta_\pi$, we have been able to
perform the phase space integrals, arriving at the analytical expression for
the decay width given by Eq.~(\ref{mainwidth}).

It is seen that the terms of the sum in Eq.~(\ref{mainwidth}), which
correspond to different combinations of hadronic form factors, are of
different orders in powers of $\eta_\pi$ [see Eqs.~(\ref{ca1a1}) to
(\ref{ca1a3})]. The dominant contribution is found to be given by the form
factor $f_{A1}$, which can be identified with the pion decay constant
$f_\pi$. With this identification, the leading order expression for the
decay width reduces to the result obtained in absence of the external field.

In fact, the expected order of the corrections arising from each combination
of form factors can be inferred from the general form of the pion-to-vacuum
amplitude [see Eqs.~(\ref{vector}) and (\ref{axial-vector})], since each
term carrying an electromagnetic strength tensor will introduce a factor
$Q\,F^{\mu\nu}\propto \xi_{\pi}\,\eta_{\pi}\,m_{\pi}^{2}$. Nevertheless, it
is found that several combinations get exactly cancelled at the leading
order. In particular, this is what happens with the contributions of order
$\eta_{\pi}$, which arise from the crossed terms $f_{A1}^{*}\,f_{V}$ and
$f_{A1}^{*}\,f_{A2}$. Due to the peculiar behavior of Bessel functions
that appear in the decay amplitude, it is found that the largest subleading
contribution to the width is the one provided by the $f_{A1}^{*}\,f_{V}$
term, which is shown to be of order $\eta_{\pi}^{4/3}$.

Our analytical results confirm previous numerical calculations obtained for
low values of the parameter $\xi_\pi$ (i.e., low values of the amplitude of
the external electromagnetic wave)~\cite{Mouslih:2020pfd}. In particular, at
leading order in powers of $\eta_\pi$, we have found analytical expressions
for the upper and lower bounds on the number of exchanged photons; these
kinematical cuts are shown to arise essentially from the properties of the
Bessel functions involved in the decay amplitude. We remark that our
expressions are valid for the full studied range of values of $\xi_\pi$. In
addition, it is shown that, at the leading order in $\eta_\pi$, our
analytical results for the contribution of a given number of exchanged
photons are analogous to those obtained from the Kroll-Watson
formula~\cite{Kroll-Watson}, which describes the cross section for the
scattering of a charged particle by some weak potential in the presence of a
strong external electromagnetic wave.

\section*{Acknowledgments}

This work has been partially funded by ANPCyT (Argentina) under Grant
No.~PICT20-01847, by CONICET (Argentina) under Grant No.~PIP 2022-2024
GI-11220210100150CO, by UNLP (Argentina) under Project No.~X960, by
Conselleria de Innovaci\'on, Universidades, Ciencia y Sociedad Digital,
Generalitat Valenciana (Spain), GVA PROMETEO/2021/083 and by Ministerio de
Ciencia, Innovaci\'on y Universidades (Spain), PID2024-158190NB-C21.

\section*{Appendices}

\appendix

\setcounter{section}{0}
\renewcommand{\thesection}{\Alph{section}}
\global\long\def\theequation{\thesection.\arabic{equation}}
\setcounter{equation}{0}
\global\long\def\thesubsection{\thesection.\arabic{subsection}}

\section{Particle fields in the presence of an electromagnetic PW background}

\label{sec:Particle-fields}

We work here under the assumption that the laser field can be described as a
PW background field. Hence, it has to be treated at all orders in
perturbation theory. This can be achieved taking into account the Furry
interaction picture, in which the background field is considered as part of
the unperturbed system~\cite{Furry:1951zz}. To obtain the particle
asymptotic states in this picture it is necessary to find the solutions of
the Klein-Gordon and Dirac equations in the presence of the background,
namely
\begin{align}
\left[\left(\partial^{\mu}+i\,Q_{\pi}A^{\mu}\right)\,
\left(\partial_{\mu}-i\,Q_{\pi}A_{\mu}\right)+m^{2}\right]\Phi^{+}(x) & \ = \ 0\ ,
\label{eq.dirac} \\[2mm]
[i\,\cancel{\partial}-Q_{l}\,\cancel{A}(x)-m]\Psi(x) & \ = \ 0 \ ,
\end{align}
where $A^\mu$ is given by Eq.~(\ref{EM_BG}). The solutions of these
equations can be found in
Ref.~\cite{Seipt:2017ckc,Mitter:1974yg,Brown:1964zzb}. To fix our notation,
in what follows we extract the main results.

\subsection{Spin 0 field}

In the presence of the electromagnetic PW background given by
Eq.~(\ref{EM_BG}), a charged spin 0 field can be expanded as
\begin{align}
\Phi^{+}(x) & \ = \ \int\,\frac{d^{3}p}{\sqrt{\left(2\pi\right)^{3}2E_{p}}}\,
\left[\,\phi_{p}^{+}(x)\,a_{\vec{p}}+\phi_{p}^{-}(x)\,b_{\vec{p}}^{\dagger}\,\right]\ ,\nonumber \\[1mm]
\Phi^{-}(x) & \ = \ \Phi^{+}(x)^{\dagger}\ = \
\int\,\frac{d^{3}p}{\sqrt{\left(2\pi\right)^{3}2E_{p}}}\,
\left[\,\phi_{p}^{+}(x)^{\ast}\,a_{\vec{p}}^{\dagger}+\phi_{p}^{-}(x)^{\ast}\,b_{\vec{p}}\,\right]\ ,
\label{pionfield}
\end{align}
where $E_{p}=\sqrt{m^{2}+\vec{p}^{\;2}}$ and $p^{\mu}=(E_{p},\vec{p}\,)$.
The wave function $\phi_{p}^{+}(x)$ is given by
\begin{equation}
\phi_{p}^{+}(x)\ =\ e^{-ip\cdot x}\,\varphi_{p}^{+}(\zeta)\ ,
\end{equation}
where
\begin{equation}
\varphi_{p}^{+}(\zeta)\ =\ \exp\left\{ -\frac{i}{k\cdot p}\int_{\zeta_{0}}^{\zeta}\,d\zeta'\,\left[Q\,A(\zeta')\cdot p-\frac{1}{2}\,Q^{2}A(\zeta')^{2}\right]\right\} \ ,\label{fi+}
\end{equation}
with $\zeta=k\cdot x$. By definition, the negative energy solution
can be obtained from the positive energy as
\begin{equation}
\phi_{p}^{-}(x)\ = \ \phi_{p}^{+}(x)^{\ast}\ ,\qquad{\rm
with\ the\ change}\quad Q\,\to\, -Q\ .
\end{equation}
In our case, this is equivalent to the transformation
$p^{\mu}\rightarrow-p^{\mu}$, for which one has
\begin{align}
\phi_{p}^{\left(-\right)}\left(x\right) \ = \ \phi_{-p}^{\left(+\right)}\left(x\right) & \ = \
\,e^{ip\cdot x/\hbar}\,\varphi_{p}^{\left(-\right)}\left(\zeta\right)\ ,
\nonumber \\
\varphi_{p}^{\left(-\right)}\left(\zeta\right) \ = \ \varphi_{-p}^{\left(+\right)}\left(\zeta\right) & \ = \
\exp\left\{ -\frac{i}{k\cdot p}\int_{\zeta_{0}}^{\zeta}\,d\zeta'\,
\left[Q\,A\left(\zeta'\right)\cdot p+\frac{1}{2}Q^{2}\,A\left(\zeta'\right)^{2}\right]\right\} \ .
\label{fi-}
\end{align}
The wave functions satisfy the orthogonality and completeness properties
\begin{align}
\int d^{4}x\;\phi_{p^{\prime}}^{(+)}(x)^\ast\,
\phi_{p}^{\left(+\right)}(x) & \ = \
(2\pi)^{4}\,\delta^{4}(p-p^{\prime}) \ , &
\int\frac{d^{4}p}{\left(2\pi\right)^{4}}\;\phi_{p}^{\left(+\right)}(x^{\prime})^\ast
\,\phi_{p}^{\left(+\right)}(x) & \ = \ \delta^{4}(x-x^{\prime})\ ,
\end{align}
where the momentum can been extended off the mass shell.

The commutation relations between creation and annihilation operators
read
\begin{align}
\Big[a_{\vec{p}},\thinspace a_{\vec{p}^{\,\prime}}^{\dagger}\Big] &
= \Big[b_{\vec{p}},\thinspace b_{\vec{p}^{\,\prime}}^{\dagger}\Big]
= \delta^{3}(\vec{p}-\vec{p}^{\,\prime})\ , \nonumber \\[1mm]
\Big[a_{\vec{p}},\thinspace a_{\vec{p}^{\,\prime}}\Big] &
= \Big[b_{\vec{p}},\thinspace b_{\vec{p}^{\,\prime}}\Big]
= \Big[a_{\vec{p}}\thinspace b_{\vec{p}^{\,\prime}}\Big]
= \Big[a_{\vec{p}},\thinspace b_{\vec{p}^{\,\prime}}^{\dagger}\Big] = 0 \ .
\end{align}
A $\pi^+$ particle state of given momentum $\vec p$ is defined by
\begin{equation}
\left|\pi^{+}\left(\vec{p}\right)\right\rangle \ = \
a_{\vec{p}}^{\dagger}\left|0\right\rangle\ .
\end{equation}

\subsection{Spin $1/2$ field}

\label{app_a_2}

A spin 1/2 field can be written as
\begin{align}
\Psi\left(x\right) & \ = \
\sum_{s=1,2}\,\int\,\frac{d^{3}p}{\sqrt{\left(2\pi\right)^{3}2E_{p}}}
\left[\boldsymbol{E}_{p}(x)\,u(\vec{p},\,s)\,c_{\vec{p},\,s}+
\boldsymbol{E}_{p}^{c}(x)\,v(\vec{p},\,s)\,d_{\vec{p},\,s}^{\dagger}\right]\ ,
\nonumber \\[1mm]
\bar{\Psi}(x) & \ = \
\sum_{s=1,2}\,\int\,\frac{d^{3}p}{\sqrt{\left(2\pi\right)^{3}2E_{p}}}
\left[\bar u(\vec{p},\,s)\,\bar{\boldsymbol{E}}_{p}(x)\,c_{\vec{p},\,s}^{\dagger}+
\bar v(\vec{p},\,s)\,\boldsymbol{\bar{E}}_{p}^{c}(x)\,d_{\vec{p},\,s}\right]\ ,
\label{FermiField}
\end{align}
where $c_{\vec{p},\,s}$ $(d_{\vec{p},\,s}^{\dagger})$ annihilates (creates)
a fermion (antifermion) of quantum numbers $\vec{p}$, $s$. The spinors
$u(\vec{p},\,s)$ and $v(\vec{p},\,s)$ are the usual free
bispinors~\cite{Seipt:2017ckc}. We use the bispinor normalization
\begin{equation}
\bar{u}(\vec{p},\,s)\,\gamma_{0}\,u(\vec{p},\,r) \ = \
\bar{v}(\vec{p},\,s)\,\gamma_{0}\,v(\vec{p},\,r) \ = \ 2E_{0}\,\delta_{sr}\ ,
\end{equation}
which corresponds to the standard normalization
\begin{equation}
\bar{u}(\vec{p},\,s)\,u(\vec{p},\,r) \ = \
-\bar{v}(\vec{p},\,s)\,v(\vec{p},\,r) \ = \ 2m\,\delta_{sr}\ .
\end{equation}
With this normalization condition one has the relations
\begin{align}
\sum_{s=1,2}\,u(\vec{p},\,s)_{\sigma}\,\bar{u}(\vec{p},\,s)_{\tau} & \ = \
(\cancel{p}+m)_{\sigma\tau} \ ,\nonumber \\[1mm]
\sum_{s=1,2}\,v(\vec{p},\,s)_{\sigma}\,\bar{v}(\vec{p},\,s)_{\tau} & \ = \
(\cancel{p}-m)_{\sigma\tau} \ .
\label{2.35}
\end{align}
The Ritus matrices in Eqs.~(\ref{FermiField}) are
\begin{align}
\boldsymbol{E}_{p}(x) & =\phi_{p}^{(+)}(x)\,\left(1+\frac{Q}{2\,k\cdot p}\,\cancel{k}\,\cancel{A}(\zeta)\right)\ ,
& \boldsymbol{E}_{p}^{c}(x) & =\phi_{p}^{(-)}(x)\,\left(1-\frac{Q}{2\,k\cdot p}\,\cancel{k}\,\cancel{A}(\zeta)\right)\ ,
\nonumber \\[2mm]
\boldsymbol{\bar{E}}_{p}(x) & = \left(1+\frac{Q}{2\,k\cdot p}\,\cancel{A}(\zeta)\,\cancel{k}\right)\,
\phi_{p}^{(+)}(x)^\ast\ ,
& \boldsymbol{\bar{E}}_{p}^{c}(x) & =
\left(1-\frac{Q\,a}{2\,k\cdot p}\,\cancel{A}(\zeta)\,\cancel{k}\right)\,\phi_{p}^{(-)}(x)^\ast\ ,
\end{align}
where have introduced the definition
$\boldsymbol{\bar{E}}=\gamma^{0}\,\boldsymbol{E}^{\dagger}\,\gamma^{0}$. In
Eqs.~(\ref{FermiField}), the negative energy solution can be obtained from
the positive energy one performing the transformations
\begin{equation}
\begin{array}{cc}
\boldsymbol{E}_{p}^{c}(x)\, =\, \mathcal{C}\,\left[\bar{\boldsymbol{E}}_{p}(x)\right]^{T}\,\mathcal{C}^{-1}\ , &
\text{with }Q\,\to\,-Q \ ,\\[2mm]
\boldsymbol{E}_{p}(x)\, =\, \mathcal{C}\,\left[\boldsymbol{\bar{E}}_{p}^{c}(x)\right]^{T}\,\mathcal{C}^{-1}\ , &
\text{with }Q\,\to\,-Q \ ,
\end{array}
\end{equation}
where
$\mathcal{C}=i\gamma^{2}\gamma^{0}=-\mathcal{C}^{\dagger}=-\mathcal{C}^{-1}$.
The transformation $p^{\mu}\rightarrow-p^{\mu}$ leads to the relations
\begin{equation}
\boldsymbol{E}_{p}^{c}(x) \ = \ \boldsymbol{E}_{-p}(x)\ ,
\qquad\qquad
\boldsymbol{\bar{E}}_{p}^{c}(x) \ = \ \boldsymbol{\bar{E}}_{-p}(x)\ .
\label{2.18}
\end{equation}
In addition, the Ritus matrices satisfy~\cite{Mitter:1974yg}
\begin{align}
\int d^{4}x\,\boldsymbol{\bar{E}}_{p}(x)\,\boldsymbol{E}_{q}(x) & \ = \
\left(2\pi\right)^{4}\delta^{4}(p-q)\ ,\qquad
\int\frac{d^{4}p}{\left(2\pi\right)^{4}}\,
\boldsymbol{\bar{E}}_{p}(x)\,\boldsymbol{E}_{p}(x^{\prime}) \ = \
\delta^{4}(x-x^{\prime})\ ,
\end{align}
where the momentum is extended off the mass shell.

The anticommutation relations for creation and annihilation operators read
\begin{align}
\Big\{ c_{\vec{p},s}\,,\,c_{\vec{p}^{\,\prime},r}^{\dagger}\Big\}  & \ = \
\Big\{ d_{\vec{p},s}\,,\,d_{\vec{p}^{\,\prime},s}^{\dagger}\Big\} \ = \
\delta^{3}(\vec{p}-\vec{p}^{\,\prime})\,\delta_{s,r} \ ,
\nonumber \\[1mm]
\Big\{ c_{\vec{p},s}\,,\,c_{\vec{p}^{\,\prime},s}\Big\}  & \ = \
\Big\{ d_{\vec{p},s}\,,\,d_{\vec{p}^{\,\prime},s}\Big\} \ = \
\Big\{ c_{\vec{p},s}\,,\,d_{\vec{p}^{\,\prime},s}\Big\} \ = \
\Big\{ c_{\vec{p},s}\,,\,d_{\vec{p}^{\,\prime},s}^{\dagger}\Big\} \ = \ 0 \ .
\end{align}
A lepton state of momentum $\vec p$ and polarization state $s$ is given by
\begin{align}
|l(\vec{p},s)\rangle  & \ = \ c_{\vec{p},s}^{\dagger}\,|0\rangle \ .
\end{align}

\section{Subleading contribution to the integrals $I_n^r$}

\label{AppB}

\setcounter{equation}{0}

In this appendix we outline the calculation of the lowest order
correction to the regularized integrals $I_n^r$ in powers of $\eta_\pi$,
Eq.~(\ref{Inreg_res}). Taking the definition of $I_n^r$ in
Eq.~(\ref{Ireg}), we start by splitting the integral into three regions,
\begin{align}
I_{n-}^{\text{r}} & =\frac{1}{\eta_{\pi}}\,\int_{\mathfrak{z}_{-}}^{\mathfrak{z}_{-}+\epsilon_{-}}\,d\mathfrak{z}\,\frac{\eta_{\pi}\,\mathfrak{z}^{n}}
{\pi\,\sqrt{\left(\mathfrak{z}-\mathfrak{z}_{-}\right)\,\left(\mathfrak{z}_{+}-\mathfrak{z}\right)+\lambda\,\eta_{\pi}^{2/3}\,\mathfrak{z}^{4/3}}}\ ,
\nonumber \\
I_{n\text{c}}^{\text{r}} &
=\frac{1}{\eta_{\pi}}\,\int_{\mathfrak{z}_{-}+\epsilon_{-}}^{\mathfrak{z}_{+}-\epsilon_{+}}\,d\mathfrak{z}\,\frac{\eta_{\pi}\,\mathfrak{z}^{n}}
{\pi\,\sqrt{\left(\mathfrak{z}-\mathfrak{z}_{-}\right)\,\left(\mathfrak{z}_{+}-\mathfrak{z}\right)+
\lambda\,\eta_{\pi}^{2/3}\,\mathfrak{z}^{4/3}}}\ ,
\nonumber \\
I_{n+}^{\text{r}} & =\frac{1}{\eta_{\pi}}\,\int_{\mathfrak{z}_{+}-\epsilon_{+}}^{\mathfrak{z}_{+}}\,d\mathfrak{z}\,\frac{\eta_{\pi}\,\mathfrak{z}^{n}}
{\pi\,\sqrt{\left(\mathfrak{z}-\mathfrak{z}_{-}\right)\,\left(\mathfrak{z}_{+}-\mathfrak{z}\right)+\lambda\,\eta_{\pi}^{2/3}\,\mathfrak{z}^{4/3}}}\ ,
\end{align}
with the assumptions
\begin{equation}
\epsilon_{\pm}\ \ll\ \left(\mathfrak{z}_{+}-\mathfrak{z}_{-}\right)
\end{equation}
and
\begin{equation}
\epsilon_{\pm}\,(\mathfrak{z}_{+}-\mathfrak{z}_{-})\ \gg\
\lambda\,\eta_{\pi}^{2/3}\,\mathfrak{z}_{\pm}^{4/3}\ .
\label{bbg}
\end{equation}
Taking into account the relation in Eq.~(\ref{bbg}), it is seen that the
integral $I_{nc}^r$ can be expanded in powers of $\eta_\pi^{2/3}$. One has
\begin{align}
I_{n\text{c}}^{\text{r}} &
=\ \int_{\mathfrak{z}_{-}+\epsilon_{-}}^{\mathfrak{z}_{+}-\epsilon_{+}}\,d\mathfrak{z}\,\frac{\mathfrak{z}^{n}}
{\pi\,\sqrt{\left(\mathfrak{z}-\mathfrak{z}_{-}\right)\,\left(\mathfrak{z}_{+}-\mathfrak{z}\right)}}\, +
\, {\cal O}(\eta_\pi^{2/3})\ ,
\end{align}
where the integral can be performed analytically. At leading order in
$\epsilon_{\pm}$ we have
\begin{align}
I_{n\text{c}}^r & =\
I_{n}-\frac{2}{\pi\,\sqrt{\mathfrak{z}_{+}-\mathfrak{z}_{-}}}\,
\left(\mathfrak{z}_{-}^{n}\,\sqrt{\epsilon_{-}}+\mathfrak{z}_{+}^{n}\,\sqrt{\epsilon_{+}}\right)\,
+ \, {\cal O}(\eta_\pi^{2/3})\ ,
\label{eqc}
\end{align}
where $I_{n}$, $n=0,1,2,$ are given in Eq.~(\ref{In_res}).

Now, to evaluate the integral $I_{n-}^{\text{r}}$ let us perform the change
\begin{equation}
\mathfrak{z}=\mathfrak{z}_{-}+\,\frac{1}{\left(\mathfrak{z}_{+}-\mathfrak{z}_{-}\right)} \; y\ .
\end{equation}
At leading order in $\epsilon_-$, one has
\begin{align}
I_{n-}^{\text{r}} \ & \simeq \ \frac{1}{\pi}\,\frac{\mathfrak{z}_{-}^{n}}{\left(\mathfrak{z}_{+}-\mathfrak{z}_{-}\right)}\,
\int_{0}^{\epsilon_{-}\left(\mathfrak{z}_{+}-\mathfrak{z}_{-}\right)}\,dy\,
\frac{1}{\sqrt{y+\lambda\,\eta_\pi^{2/3}\,\mathfrak{z}_{-}^{4/3}}}\nonumber \\
 & \simeq\ \frac{2}{\pi}\,\mathfrak{z}_{-}^{n}
 \left[ \frac{\sqrt{\epsilon_{-}}}{\sqrt{\mathfrak{z}_{+}-\mathfrak{z}_{-}}}\,
 -\,\sqrt{\lambda}\,\frac{\mathfrak{z}_{-}^{2/3}}{\left(\mathfrak{z}_{+}-\mathfrak{z}_{-}\right)}\,\eta_{\pi}^{1/3}\right]
 \, + \, {\cal O}(\eta_{\pi}^{2/3})\ .
\label{eqmen}
\end{align}
In the same way, for the integral $I_{n+}^{\text{r}}$ we can perform
the change $\mathfrak{z}=\mathfrak{z}_{+}-y/(\mathfrak{z}_{+}-\mathfrak{z}_{-})$. This leads to
\begin{align}
I_{n+}^{\text{r}} \ & \simeq\ \frac{2}{\pi}\,\mathfrak{z}_{+}^{n}
 \left[ \frac{\sqrt{\epsilon_{+}}}{\sqrt{\mathfrak{z}_{+}-\mathfrak{z}_{-}}}\,
 -\,\sqrt{\lambda}\,\frac{\mathfrak{z}_{+}^{2/3}}{\left(\mathfrak{z}_{+}-\mathfrak{z}_{-}\right)}\,\eta_{\pi}^{1/3}\right]
 \, + \, {\cal O}(\eta_{\pi}^{2/3})\ .
\label{eqmas}
\end{align}
Summing the expressions in Eqs.~(\ref{eqc}), (\ref{eqmen}) and (\ref{eqmas})
the dependences on $\epsilon_\pm$ cancel (as they should do), leading to the
result in Eq.~(\ref{Inreg_res}).

\end{document}